\documentclass{jfmX}

\usepackage{graphicx}
\usepackage{natbib}
\usepackage{amssymb,amsmath}
\usepackage{color}
\input colors.defs

\newcommand{\bminil}[1]{\begin{minipage}[l]{#1 \textwidth}}
\newcommand{\bminir}[1]{\begin{minipage}[r]{#1 \textwidth}}
\newcommand{\bminic}[1]{\begin{minipage}[c]{#1 \textwidth}}
\newcommand{\emini}{\end{minipage}}
\newcommand{\EQL}{\begin{equation}\label}
\newcommand{\EQ}{\begin{equation}}
\newcommand{\EN}{\end{equation}}
\newcommand{\BFG}{\begin{figure}}
\newcommand{\EFG}{\end{figure}}
\newcommand{\ITM}{\begin{itemize}}
\newcommand{\ITN}{\end{itemize}}
\newcommand{\ENM}{\begin{enumerate}}
\newcommand{\EEN}{\end{enumerate}}
\newcommand{\BEA}{\[\begin{array}}
\newcommand{\EEA}{\end{array}\]}
\newcommand{\EQAL}{\begin{eqnarray}\label}
\newcommand{\EQA}{\begin{eqnarray}}
\newcommand{\ENA}{\end{eqnarray}}

\newcommand{\btriangle}{\mbox{\boldmath$\triangle$}}

\newcommand{\bdot}{\mbox{\boldmath$\cdot$}}

\newcommand{\bB}{\mbox{\boldmath$B$}}

\newcommand{\bN}{\mbox{\boldmath$N$}}

\newcommand{\bS}{\mbox{\boldmath$S$}}

\newcommand{\bff}{\mbox{\boldmath$f$}}

\newcommand{\bj}{\mbox{\boldmath$j$}}
\newcommand{\bk}{\mbox{\boldmath$k$}}

\newcommand{\br}{\mbox{\boldmath$r$}}

\newcommand{\bu}{\mbox{\boldmath$u$}}

\newcommand{\bx}{\mbox{\boldmath$x$}}
\newcommand{\by}{\mbox{\boldmath$y$}}

\newcommand{\bGamma}{\mbox{\boldmath$\Gamma$}}

\newcommand{\bomega}{\mbox{\boldmath$\omega$}}

\newcommand{\p}{\partial}

\newcommand{\ppto}[1]{\frac{\partial #1}{\partial t}}

\newcommand{\half}{\mbox{$\frac{1}{2}$}}

\newcommand{\ddto}[1]{\frac{d #1}{dt}}

\renewcommand{\theenumi}{\alph{enumi}}

\newcommand{\bbR}{\mathbb{R}}
\newcommand{\bbT}{\mathbb{T}}

\newcommand{\bbZ}{\mathbb{Z}}

\newcommand{\sbl}[3]{\|{#1}\|_{#2}^{#3}} 

\newcommand{\Lap}{\mbox{\boldmath$\triangle$}}
 
\def\DEL#1{\textcolor{Green}{}}     
\def\AD#1{{\textcolor{magenta}{}}}       
\def\RoraiQ#1{{\textcolor{cyan}{}}}         
\def\RMKA#1{}         
\def\xref#1{}         

\ifCUPmtlplainloaded \else
  \checkfont{eurm10}
  \iffontfound
    \IfFileExists{upmath.sty}
      {\typeout{^^JFound AMS Euler Roman fonts on the system,
                   using the 'upmath' package.^^J}%
       \usepackage{upmath}}
      {\typeout{^^JFound AMS Euler Roman fonts on the system, but you
                   dont seem to have the}%
       \typeout{'upmath' package installed. JFM.cls can take advantage
                 of these fonts,^^Jif you use 'upmath' package.^^J}%
      }
  \else
  \fi
\fi

\ifCUPmtlplainloaded \else
  \checkfont{msam10}
  \iffontfound
    \IfFileExists{amssymb.sty}
      {\typeout{^^JFound AMS Symbol fonts on the system, using the
                'amssymb' package.^^J}%
       \usepackage{amssymb}%
         \let\leq=\leqslant
         \let\geq=\geqslant
      }{}
  \fi
\fi

\ifCUPmtlplainloaded \else
  \IfFileExists{amsbsy.sty}
    {\typeout{^^JFound the 'amsbsy' package on the system, using it.^^J}%
     \usepackage{amsbsy}}
    {}
\fi

\newcommand\etall{\mbox{\textit{et al.}}}

\newcommand{\emailnotes}[1]{}
\newcommand{\biband}{\&~}
\newcommand{\authone}[2]{#2,~#1}
\newcommand{\authtwo}[4]{#2,~#1,~\&~#4,~#3}
\newcommand{\auththr}[6]{#2,~#1,~#4,~#3,~\&~#6,~#5}
\newcommand{\authfour}[8]{#2,~#1,~#4,~#3,~#6,~#5,~\&~#8,~#7} 
\newcommand{\authmanytwo}[4]{#2,~#1,~#4,~#3,}

\newcommand{\yjour}[6]{ #1~ #6. {\em #2} {\bf #3}, #4#5.}

\newcommand{\yweb}[3]{ #1~ #2 #3.}

\title[Scaling of Navier-Stokes trefoil reconnection]{Scaling of Navier-Stokes trefoil reconnection}

\author[R. M. Kerr]
{Robert M. Kerr
  \thanks{Email address for correspondence: Robert.Kerr@warwick.ac.uk},\ns}

\affiliation{Department of Mathematics, University of Warwick,
Coventry CV4 7AL, United Kingdom}

\pubyear{}
\volume{}
\pagerange{}
\begin{document}

\maketitle

\begin{abstract} 
Perturbed, helical trefoil vortex knots and a set of anti-parallel vortices are examined 
numerically to identify the scaling of their helicity and vorticity norms during 
reconnection.  For the volume-integrated enstrophy $Z=\int_{\bbT_\ell^3}\omega^2 dV$,
a new scaling regime is identified for both configurations where as the viscosity $\nu$
changes, all $\sqrt{\nu}Z(t)$ cross at $\nu$-independent times $t_x$,
identified as when the first reconnection events end. 
Self-similar linear collapse of $B_\nu(t)=(\sqrt{\nu}Z)^{-1/2}$ can be found for
$t\lesssim t_x$ by linearly extrapolating $B_\nu(t)$ to zero at critical times 
$T_c(\nu)$, then plotting $(T_c(\nu)-t_x)\bigl(B_\nu(t)-B_x\bigr)$ where $B_x=B_\nu(t_x)$.
The size $\ell^3$ of the periodic domains must be increased as $\nu$ is decreased 
to maintain this scaling as implied by known Sobolev space bounds. 
The anti-parallel calculations show that the linear collapse of $B_\nu(t)$ 
begins with a quick, viscosity-independent exchange of the circulation $\Gamma$ between 
the original vortices and the new vortices.  
Up to and after the trefoil knots' first reconnection at time $t_x$,
their helicity ${\cal H}$ is preserved, validating the experimental centreline helicity 
observation of \cite{ScheeleretalIrvine2014a}. 
Because the cubic Navier-Stokes velocity norm $\sbl{u}{L^3}{}$ barely 
changes and the Navier-Stokes $\|\omega\|_\infty$ are bounded by the Euler values,
these flows are never singular. Despite this, the Navier-Stokes $Z$ can, 
for a brief period, grow faster than the Euler $Z$ 
and the following increase in the viscous energy 
dissipation rate $\epsilon=\nu Z$ shows $\nu$-independent convergence at $t\approx 2t_x$. 
Taken together, these results could be a new paradigm whereby smooth solutions 
without singularities or roughness could generate a $\nu\to0$ 
{\it dissipation anomaly} (finite dissipation in a finite time) as $\ell\to\infty$,
as seen in physical turbulent flows.  
\end{abstract}

\newpage
\section{Background}
Two long-standing questions about nonlinear growth in fluid turbulence are these:
Does hydrodynamic helicity have a role in controlling nonlinearity that is analogous to the role
of the helicity of magnetic systems \citep{Moffatt2014}? And 
can there be smooth nonlinear growth whose scaling is consistent with the formation 
of a Navier-Stokes {\it dissipation anomaly}?
That is, can there be finite energy dissipation in a finite time as the viscosity $\nu\to0$?
The first question needs revisiting due to the recent vortex knot reconnection experiments of 
\cite{ScheeleretalIrvine2014a} that observed complete reconnections, indicating
strong nonlinearities, while simultaneously retaining strong helicity that could have 
suppressed that growth.  Inspired by those experiments, this paper gains insight into 
both questions using new high-resolution simulations of reconnecting trefoil vortex knots and 
anti-paralllel vortices. 

The {\it dissipation anomaly} question
is addressed by considering the following dichotomy. \cite{JCV15} quotes 
several sources to conclude that there is a consensus that whenever an energy cascade is observed, 
the energy dissipation rate $\epsilon=\nu Z$ is independent of $\nu$,  
where $Z(t)$ is the volume-integrated enstrophy \eqref{eq:enstrophy}.  This
contrasts with the mathematics that shows that unless there are 
singularities, $\nu\to0$ finite dissipation cannot form when the domain
is fixed \citep{Constantin86}.  Could the unexpected preservation of helicity 
during the reconnection of the experimental trefoil knot of
\cite{ScheeleretalIrvine2014a} provide clues to resolving this dichotomy? 

\begin{figure} 
\includegraphics[scale=0.32]{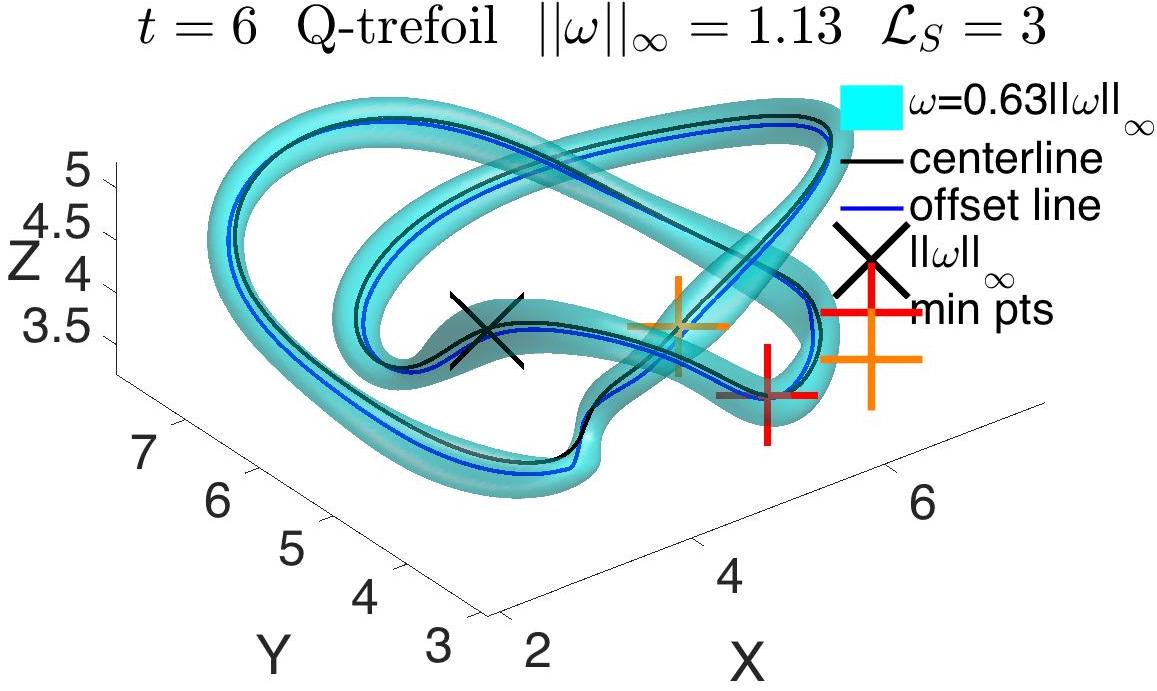}
\caption{\label{fig:T6} Vorticity isosurface plus two closed vortex lines
of the perturbed trefoil vortex at $t=6$, not long after initialisation. 
One vortex line has been seeded at $\|\omega\|_\infty$ and the other a bit offset, allowing
the self-linking ${\cal L}_S$ to be calculated using \eqref{eq:Gauss} giving
${\cal L}_S=3$, which can be split \eqref{eq:selflink} into 
a writhe of $W\!r=3.15$ and a twist of $T\!w=-0.15$. {\bf X} is the location of $\|\omega\|_\infty$
and the seed for the black vortex line. The red and yellow {\bf +} are the points of
closest approach of the two loops of the trefoil and where reconnection will begin.} 
\end{figure}

The trefoils being simulated have the topology of a (2,3) knot, are inherently helical and evolve 
as self-reconnecting, doubly-looped rings.  As described in section \ref{sec:initialisation}, 
to properly reproduce the single, dominant reconnection event of the experimental trefoils, 
perturbations are needed to break the three-fold symmetry of an ideal trefoil. 
The resulting initial state is given in figure
\ref{fig:T6} and once this perturbed state was generated, the first task was 
to establish the range of viscosities and thicknesses that could be run using comparisons of
peak vorticities $\|\omega\|_\infty$ for different resolutions and domains
as in figures \ref{fig:EulerNSomaxtsmall} and evolving enstrophy spectra $Z(k,t)$ as in figure
\ref{fig:NSomspectra}. 
This initial analysis also provided preliminary trends for how trefoil reconnnection events 
can be suppressed, enhanced and develop signs of self-similarity, 

The next task would have been to examine whether the \cite{ScheeleretalIrvine2014a} claim that 
helicity can be preserved during the first reconnection, which is now confirmed 
in figure \ref{fig:HL3H12},

However, the helicity analysis was soon overshadowed by the identification of new self-similar 
scaling regime for the evolution of the volume-integrated enstrophy $Z(t)$ during the first 
reconnection.  The kernel of this scaling regime is using $\sqrt{\nu}Z(t)$, $\nu$ viscosity,
instead of the expected dissipation rate scaling of $\epsilon(t)=\nu Z(t)$, and
was first identified using figure \ref{fig:QsnuZ}a. After some exploration it was found to
manifests itself more completely in figure \ref{fig:QHdRisnuZtime} as a
viscosity-independent, linear collapse of $(\sqrt{\nu}Z(t))^{-1/2}$ until the first reconnection 
ends at $t=t_x$. This scaling has probably not been noticed because it requires increasing the
size of the domain $\ell$ as $\nu\to0$ due to mathematical bounds upon the growth of higher-order 
norms in finite periodic domains most of the fluids community is not familiar with. This
behaviour could apply to all strong reconnection events as 
the collapse is also found for the new anti-parallel vortex reconnection calculations in
figure \ref{fig:dNSsnuZ}. Its importance is that this scaling regime
could be a precursor to the formation of a $\nu\to0$, $\ell\to\infty$ 
{\it dissipation anomaly} from smooth solutions at later times, as demonstrated by the
energy dissipation rates $\epsilon(t)=\nu Z(t)$ plotted in figure \ref{fig:QsnuZ}b. 

It is important to place these two surprising results, helicity preservation 
and the new $\sqrt{\nu}Z(t)$ enstrophy scaling, in the context of what is already known from
simulations about the role of helicity and known results from applied analysis of the 
Navier-Stokes about what could potentially bound the growth of the enstrophy $Z(t)$.

If the initial state is a collection of random large-scale helical Fourier modes, simulations
show that the onset of turbulence can be delayed. However, the subsequent decay is 
similar to what non-helical initial conditions generate \citep{Polifkeetal1989}. 
Another Fourier approach is to insert anisotropic modes or forcing at intermediate 
wavenumbers to determine the influence of the sign of helicity upon the energy cascade
\citep{BiferaleKerr95,Sahooetal2015}.  
Helicity is also a continuum measure of the topology of the vortices \citep{MoffattRicca92},
so can serve as a diagnostic for changes in the topology.  However, up to now this use 
of helicity has been limited because the configurations that are easy to construct, 
such as anti-parallel and orthogonal pairs, have zero or weak helicity.  
The only calculations with some helicity have used skewed initial states, 
either random \citep{HolmKerr07} or vortical \citep{KimuraMoffatt2014}.  These have 
reinforced the perspective that helicity tends to suppress nonlinearity before it
disappears. This question will be revisited in section \ref{sec:Hnegative} with 
respect to the generation of negative helicity.

Based upon that experience, the experimental observation that a trefoil's helicity 
can be preserved during topology changing reconnections was unexpected. 
Instead, it seems that the trefoil's self-linking helicity 
can be converted directly into the mutual helicity of new linked rings during reconnection,  
as suggested by \cite{Laingetal2015}. 

Regularity questions and results arise due to the $\nu\to0$, viscosity-independent 
convergence of the scaled enstrophy $\sqrt{\nu}Z(t)$ at $t=t_x\approx40$ in 
figure \ref{fig:QsnuZ} and the circulation exchange $\epsilon_\Gamma(t)$ \eqref{eq:dGammadt} 
at $t=t_\Gamma\approx16.5$ in figure \ref{fig:GammasepsG}. These imply respectively that 
$Z(t_x)\sim \nu^{-1/2}$ \eqref{eq:snuZLeray} and that the velocity second-derivatives 
\eqref{eq:ppu} in $\epsilon_\Gamma$ go as $\nu^{-3/2}$.  
In contrast to these trends for growth as $\nu\to0$, the two diagnostics primarily used for showing
regularity of the Navier-Stokes equation as defined by the Clay prize problem \citep{Clay}, 
imply regularity for all time. 
One diagnostic is the Navier-Stokes vorticity maximum $\|\omega\|_\infty$ \eqref{eq:ominfty}, 
whose growth is bounded by the regular Euler values in figure \ref{fig:EulerNSomaxtsmall}. 
The other diagnostic is the cubic velocity norm $\sbl{u}{L^3}{}$ \eqref{eq:LpV}. By
\cite{EscauSS03}, there can be singularities of Navier-Stokes only if $\sbl{u}{L^3}{}$ is singular 
and instead, figure \ref{fig:HL3H12}c shows that $\sbl{u}{L^3}{}$ is nearly independent 
of the viscosity and time. 
Is there mathematics that might be consistent with these opposing trends?

The mathematical result most relevant to these calculations is a proof for how regular, 
non-singular Euler solutions can bound Navier-Stokes solutions in fixed domains $\ell^3$ 
as $\nu\to0$ \citep{Constantin86}. What was shown is that so long as
solutions under Euler are bounded, critical viscosities $\nu_s$ can be
found for each $\sbl{u}{H^s}{}$ norm such that, for any
$\nu<\nu_s$, the Navier-Stokes solutions will be bounded by functions of the regular Euler 
solutions. This result is largely unknown in the fluids community and comes close to saying 
that a $\nu\to0$ {\it dissipation anomaly} cannot form. This is because in fixed domains
these bounds would preclude the viscosity-independent convergence of both $\sqrt{\nu}Z(t)$ 
and $\epsilon=\nu Z(t)$ in figure \ref{fig:QsnuZ}.  Section \ref{sec:increasingell} 
will show how this constraint can be relaxed by increasing $\ell$ as $\nu$ decreases, 
potentially allowing a $\nu\to0$ {\it dissipation anomaly} to form. 
The remaining mathematical caveats upon that statement are 
discussed in sections \ref{sec:regularity} and \ref{sec:summary}.

Is the trefoil an appropriate configuration for investigating fundamental questions about 
the regularity of the Navier-Stokes equations?  In fact,
the trefoil is well-suited for these questions for the following reasons.
First, a trefoil self-reconnects.  Second, unlike other configurations such 
as initially anti-parallel or orthogonal vortices, most of a trefoil's velocity and
vorticity norms are finite in an infinite domain, including the
enstrophy $Z$ \eqref{eq:enstrophy} and the helicity ${\cal H}$ \eqref{eq:helicity}.
Third, opposing the tendency to self-reconnect, the trefoil knots can be used to investigate 
how helicity suppresses nonlinearities because the initial helicity 
is unusually close to the upper bound defined by the energy and enstrophy, 
${\cal H}\leq(2EZ)^{1/2}$. 
Fourth, it can be simulated in a periodic box, making detailed Fourier analysis 
using Sobolev norms possible.

This paper is organised as follows. After introducing the equations, diagnostics and 
initialisation, illustrated for the Q-trefoil in figure \ref{fig:T6}, there is a step-by-step 
description of how the $\sqrt{\nu}Z(t)$ scaling for the Q-trefoil in figure \ref{fig:QsnuZ}a 
can be transformed into a viscosity-independent, linear collapse of $(\sqrt{\nu}Z(t))^{-1/2}$ 
up to a time of $t=t_x\approx40$ in figure \ref{fig:QHdRisnuZtime}, including the mathematics that 
underlies why the domain size $\ell$ must be increased as the viscosity $\nu$ is decreased. 
This is the first major result and is not limited to trefoils, as shown by new anti-parallel 
calculations where this scaling begins with a spurt of circulation exchange and
ends when the first reconnection completes.
For the trefoils, vorticity isosurfaces at $t=31$ and $t=45$ also show that $t_x\approx40$ is 
when its first reconnection ends. 
Later in time, for $t>t_x$, the trefoil's enstrophy growth increases until a finite, 
$\nu$-independent dissipation rate $\epsilon=\nu Z$ forms at $t\approx2t_x$ in a manner
consistent with the formation of a {\it dissipation anomaly}. 
The second major result is confirmation of the experimental preservation of the trefoil helicity 
${\cal H}$ to about $1.5t_x$, after the first reconnection ends, including cases
with a core radius similar to the latest experiments \citep{ScheeleretalIrvine2014a}.
The two regularity diagnostics with the same dimensions as ${\cal H}^{1/2}$, 
$\sbl{u}{L^3(\bbT_\ell^3)}{}$ and $\sbl{u}{\dot{H}^{1/2}(\bbT_\ell^3)}{}$, 
are also plotted and discussed. 
Finally, figure \ref{fig:EulerNS-Zttx} shows that the self-similar scaling of the growth of
enstrophy $Z$ continues to extremely small viscosities despite the empirical evidence 
in figure \ref{fig:EulerNSomaxtsmall}
that $\|\omega\|_\infty$ is bounded by the regular Euler solutions.

\subsection{Equations and continuum diagnostics \label{sec:equations}}

The governing equations for the simulations in this paper will be the incompressible 
Navier-Stokes equations on the toris $\bbT_\ell^3$, that is a periodic box with volume
($V=\ell^3$). 
\EQL{eq:NS} \begin{array}{r}\bu_t-\nu\Lap\bu+(\bu\bdot\nabla)\bu
+\nabla p=0\\ \nabla\bdot\bu=0 \end{array}\Bigr\}\quad 
{\rm in}~ \bbR^3\times(-\infty,T) \EN
The equations for the densities of the energy, enstrophy and helicity,
$e=\half|\bu|^2$, $|\bomega|^2$ and $h=\bu\cdot\bomega$ respectively are (with their 
volume-integrated norms):
\EQL{eq:energy} \ppto{e}+ ({\bu}\cdot\nabla)e = -\nabla\cdot(\bu p) 
+\nu\triangle e
-\underbrace{\nu(\nabla\bu)^2}_{\epsilon={\rm dissipation}},\qquad 
E=\half\int\bu^2dV\,.\EN
\EQL{eq:enstrophy} \ppto{|\bomega|^2}+ ({\bu}\cdot\nabla)|\bomega|^2 = 
\underbrace{2\bomega\bS\bomega}_{Z_p={\rm production}}
+\nu\triangle|\bomega|^2
-  \underbrace{2\nu(\nabla\bomega)^2}_{\epsilon_\omega=Z-{\rm dissipation}},\qquad 
Z=\int\bomega^2dV\,.\EN
\EQL{eq:helicity} \ppto{h}+ ({\bu}\cdot\nabla)h = 
\underbrace{-\bomega\cdot\nabla\Pi}_{\omega-{\rm transport}}
+\underbrace{\nu\btriangle h}_{\nu-{\rm transport}} -\underbrace{
2\nu{\rm tr}(\nabla\bomega\cdot\nabla\bu^T)}_{\epsilon_h={\cal H}-{\rm dissipation}}
\qquad
{\cal H}=\int\bu\cdot\bomega dV\,,\EN
where $\bomega=\nabla\times\bu$ is the vorticity vector and $\Pi=p-\half\bu^2\neq p_h$
(pressure head $p_h=p+\half\bu^2$).

The numerical method will be a 2/3rds-dealiased pseudo-spectral code with a very high 
wavenumber cut-off filter \citep{Kerr2013a}.

The inviscid ($\nu=0$) invariants are the global energy $E$ and helicity ${\cal H}$ 
\citep{Moffatt1969}, plus the circulations $\Gamma_i$ about the vortices 
\EQL{eq:Gamma} \Gamma_i=\oint \bu_i\cdot d\br_i \quad{\rm where}\quad \br_i~~
\mbox{is a closed loop about trajectories}~~{\bx_i\in\cal C}_i(t)\,. \EN
The initial trajectories
$\bx_i\!\in\!{\cal C}_i(0)$ were used to map vorticity onto the mesh and for 
the $\nu\neq0$ anti-parallel reconnection, figure \ref{fig:GammasepsG} uses
the symmetry plane diagnostics $\Gamma_y(t)$, $\Gamma_z(t)$ \eqref{eq:Gammayz} and
$\epsilon_\Gamma$ \eqref{eq:dGammadt} to show the scaling of the reconnection as it begins.

Since the energy in all of the simulations barely decays and the circulation of the
trefoil vortex is difficult to determine, changes in the global helicity ${\cal H}$, which
has the same dimensional units as the circulation-squared and can be of either sign,
has a greater role in characterising the temporal evolution.  
Furthermore, the sign of the local helicity density $h$ can grow, decrease 
and even change due to both the $\omega$-transport term along the vortices and the viscous terms
in \eqref{eq:helicity} \citep{BiferaleKerr95}. 
The evolution of the enstrophy $Z$ \eqref{eq:enstrophy} is simpler as
its production term $Z_p$ is predominantly positive and its dissipation term is strictly
negative-definite.

In addition to $E$, ${\cal H}$ and $Z$, the characterisation of the temporal evolution will
use a few volume-integrated versions of the Lebesgue measures and Sobolev-$\dot{H}$ norms.
Volume-integrated norms $\bbT_\ell^3$ of the periodic $\ell^3$ computational domains 
are used instead of the more common volume-averaged versions 
so that calculations using different $\ell$ can be compared to each other
and to mathematical results for whole space, that is in $\bbR^3=\bbT^3_\infty$.
The relationships between the volume-integrated $\sbl{u}{L^p(\bbT_\ell^3)}{}$ 
and the volume-averaged $\sbl{u}{L^p}{}$ Lebesgue measures are:
\EQL{eq:LpV} 
\sbl{u}{L^p(\bbT_\ell^3)}{}= \left(\int_{\bbT_\ell^3} d\Omega |\bu|^p\right)^{1/p}=
\ell^{3/p} \left(\frac{1}{V}\int_{V=\bbT_\ell^3} d\Omega |\bu|^p\right)^{1/p}
=\ell^{3/p} \sbl{u}{L^p}{} \,. \EN
Using $\hat{\bu}(\bk)$ as the Fourier transform of $\bu(\bx)$, 
the volume-integrated Sobolev-${\dot{H}}^s$ norms $\sbl{u}{\dot{H}^s(\bbT_\ell^3)}{}$
are summed over the $\bk=\bj\Delta k$, where the $\bj\!\in\!\bbZ^3$ are the 3D integers 
and $\Delta k=2\pi/\ell$.  The correspondence between the $\sbl{u}{\dot{H}^s(\bbT_\ell^3)}{}$
and the averaged $H^s$ norms is:
\EQL{eq:HsV} 
\sbl{u}{\dot{H}^s(\bbT_\ell^3)}{}= 
\left(\sum_{\bj} \frac{|\bk|^{2s}|}{\Delta k^3}\hat{\bu}(\bk)|^2\right)^{1/2}\!
=\left(\frac{\ell}{2\pi}\right)^{3/2} \left(\sum_{\bj} |\bk|^{2s}|\hat{\bu}(\bk)|^2\right)^{1/2}\!
=\left(\frac{\ell}{2\pi}\right)^{3/2} \sbl{u}{\dot{H}^s}{} \,.  \EN

The standard $H^s$ Sobolev-norms with a pre-factor of $1+k^2+\dots+k^{2s}$ are not being used,
except when referencing literature that uses the $H^s$ norms. To be dimensionally consistent
this would have to be a sum of $(k/k_0)^m$ with $k_0\sim\ell^{-1}$.

The relationships between the periodic volume-integrated norms on $\bbT_\ell^3$ 
as $\ell\to\infty$ and the $\bbR^3$ norms over whole space are
\EQL{eq:LpHsT3R3} 
\lim_{\ell\to\infty}\sbl{u}{L^p(\bbT_\ell^3)}{}=\sbl{u}{L^p(\bbR^3)}{}=
\left(\int_{\bbR^3} d\Omega |\bu|^p\right)^{1/p}\quad{\rm with}\quad 
L^{(3)}_\ell=\sbl{u}{L^3(\bbT^3_\ell)}{} \EN
and
\EQL{eq:LpHsT3R3b} 
\lim_{\ell\to\infty}\sbl{u}{\dot{H}^s(\bbT_\ell^3)}{}=
\sbl{u}{\dot{H}^s(\bbR^3_k)}{}
=\left(\int_{\bbR^3_k} d^3k |\bk|^{2s}|\hat{\bu}(\bk)|^2\right)^{1/2}\quad{\rm with}\quad  
\dot{H}^{(1/2)}_\ell=\sbl{u}{\dot{H}^{1/2}(\bbT_\ell^3)}{} \,.
\EN

Besides the volume-integrated enstrophy $Z=\sbl{\omega}{L^0_\ell}{2}=\sbl{u}{\dot{H}^1}{2}$ 
(\ref{eq:enstrophy},\ref{eq:LpV},\ref{eq:HsV}), the other significant 
vorticity diagnostic will be the maximum of vorticity 
\EQL{eq:ominfty} \|\bomega\|_\infty=\|\bomega\|_{L^\infty}=\sup|\bomega|\,. \EN
Its time integral will bound any property \citep{BKM84} in the sense that:
\EQL{eq:BKM} {\rm Given}\quad I(T)=\int_0^T \|\omega\|_\infty dt < \infty\quad{\rm then}\quad
\sbl{u(T)}{L^p}{}\leq \sbl{u(0)}{L^p}{}\exp\bigl(C_p(\ell) I(T)\bigr) \,. \EN
However, this leaves open what the $\ell$ dependence of $C_p$ is as $\ell$ increases.  
For example, if one takes the 
standard H\"older inequality $\sbl{\omega}{L^p}{}\leq \tilde{C}_p\sbl{\omega}{L^{p+1}}{}$, 
with $\tilde{C}_p$ independent of $\ell$, 
then by \eqref{eq:LpV} and using induction to form a constant $\tilde{C}_{2\infty}$, 
the volume-integrated enstrophy $Z$ is bounded from above as
\EQL{eq:Holder} Z=\sbl{\omega}{L^2(\bbT_\ell^3)}{2}=\ell^3\sbl{\omega}{L^2}{2}{}
\leq \tilde{C}_{2\infty}\ell^3\sbl{\omega}{\infty}{2}\,, \EN
which includes the volume as $\ell^3$.

The behaviour of the scaled helicity ${\cal H}^{(1/2)}$ \eqref{eq:helicity} and its two 
partners, the cubic velocity norm $L^{(3)}_\ell$ \eqref{eq:LpHsT3R3} and 
$\dot{H}^{(1/2)}_\ell$ \eqref{eq:LpHsT3R3b} are given in figure \ref{fig:HL3H12}.
All three of these diagnostics 
have the same units as the circulation: $[\Gamma]=[L^2/T]$ and both $L^{(3)}_\ell$ and
$\dot{H}^{(1/2)}_\ell$ have been used to improve
our understanding of the regularity of the Navier-Stokes equations 
\citep{EscauSS03,Doering09,Seregin2011}. Section \ref{sec:Leray} on applications of \cite{Leray34}
scaling discusses $L^{(3)}_\ell$ which is currently the most
refined upper bound on Navier-Stokes regularity, along with a possible origin for the 
$\sqrt{\nu}Z(t)$ scaling.
The  norm $\dot{H}^{(1/2)}_\ell$ is discussed at the end of section \ref{sec:helicity}.

\subsection{Vortex lines and linking numbers\label{sec:link}}

The role of the helicity ${\cal H}$ in understanding the trefoil calculations is two-fold. 
One role is as a constraint upon the growth of nonlinearity in the continuum, 
the other is to provide a qualitative picture of 
how the topology can change using the linking numbers. For the continuum equations, ${\cal H}$ 
is the volume-integral of the helicity density $h=\bu\cdot\bomega$ and is conserved by the 
inviscid equations, with $h$ governed by \eqref{eq:helicity}. 
In a three-dimensional turbulent flow, the kinetic energy 
cascades overwhelmingly to small scales. In contrast, $h$ can move to both large and 
small scales. 

While the global helicity ${\cal H}$ could in principle be estimated topologically
by summing contributions from selected vortex trajectories,
in this paper only a few of these trajectories are identified. The topological properties 
of these trajectories are determined to provide a qualitative picture of the changes to
the vortex structures during reconnection that can be used for determining timescales for comparisons
with the trefoil experiments \citep{KlecknerIrvine2013,ScheeleretalIrvine2014a}.

When the vortices are distinct, this topological helicity ${\cal H_L}$ can in principle
be formed using the circulations $\Gamma_i$ about the vortex trajectories 
${\cal C}_i(t)$ \eqref{eq:Gamma}, 
the linking numbers ${\cal L}_{ij}$ between all vortices and the integer self-linking numbers 
of individual closed loops 
\EQL{eq:selflink} {\cal L}_{Si}=W\!r_i+T\!w_i\,, \EN
where ${\cal L}_{Si}$ is a sum of the non-integer writhe and twist 
$W\!r_i$ and $T\!w_i$ \citep{MoffattRicca92}.  By assigning
circulations $\Gamma_i$ to the vortices and summing one can determine the global helicity
${\cal H}$ \citep{MoffattRicca92}. 
\EQL{eq:Hlink} {\cal H_L}=\sum_{ij}\Gamma_i\Gamma_j{\cal L}_{ij} +\sum_{i}\Gamma_i^2 {\cal L}_{S_i}
\,.\EN
\DEL{The four topological numbers just introduced are: 
The intervortex linking numbers ${\cal L}_{ij}$ between 
distinct vortex trajectories $\bx_i\in{\cal C}_i$ and $\bx_j\in{\cal C}_j$. The
writhe $Wr_i$ and twist $Tw_i$ numbers of a given vortex. And their
sum, the self-linking number:  
\EQL{eq:selflink} {\cal L}_{Si}=Wr_i+Tw_i\,.\EN 
The ${\cal L}_{ij}$ and ${\cal L}_{Si}$ for closed loops will be integers 
\citep{Pohl68}, but that is not a requirement for either the writhe
$Wr$ or the twist $Tw$. }

The tool used to determine the writhe, direct self-linking and intervortex linking 
for the selected vortex lines in figures \ref{fig:T6}, \ref{fig:T31} and \ref{fig:T45} 
is a regularised Gauss linking integral 
about two loops $\bx_{i,j}\in{\cal C}_{i,j}$
\EQL{eq:Gauss} {\cal L}_{ij} =
\sum_{ij}\frac{1}{4\pi}\oint_{{\cal C}_i}\oint_{{\cal C}_j}
\frac{(d\bx_i\times d\bx_j)\cdot(\bx_i-\bx_j)}{(|\bx_i-\bx_j|^2+\delta^2)^{1.5}}\,. 
\EN
The regularisation of the denominator using $\delta$ has been added for determining the
writhe when $i=j$ \citep{Calugareanu59,MoffattRicca92}. 
The self-linking numbers ${\cal L}_{Si}$ can be determined directly using $\delta=0$
by defining the edges of vortex ribbons from two parallel trajectories within 
the vortex cores, as illustrated in figure \ref{fig:T6}. 
For determining the intervortex linking numbers with $i\neq j$, $\delta=0$.  
\DEL{\indent The intrinsic twist of a closed loop $Tw_i$ can be 
determined from the line integral of the torsion of the vortex lines:
\EQL{eq:twist} Tw_i=\frac{1}{2\pi}\oint \tau ds,\qquad{\rm where}\quad
\tau=\frac{d\bN}{ds}\cdot\bB\,,\EN
where the curvature $\kappa$ and torsion $\tau$ are determined by the Frenet-Serret 
relations of a curve in space: $\bx(s):[0,1]\rightarrow\bbR^3$. A complete discussion of
the linking numbers for one of these calculations will be the topic of another paper.}

To determine these linking numbers and provide qualitative comparisons with the 
experimental vortex lines \citep{KlecknerIrvine2013,ScheeleretalIrvine2014a},
vortex lines $\bx_j(s)$ were identified by solving the 
following ordinary differential equation, 
\EQL{eq:vortexlines} \frac{d\bx_j(s)}{ds}=\bomega(\bx_j(s))\,. \EN
This can be easily solved using the Matlab streamline function, which includes a function for
interpolating the vorticity vector field $\bomega(\bx_j)$ from the mesh, onto the lines. The
seeds for solving \eqref{eq:vortexlines} were chosen from the positions around, but not 
necessarily at, local vorticity maxima. 
\section{Initialisation and domain\label{sec:initialisation}}

\begin{table}
  \begin{center}
\begin{tabular}{cccllccccccc}
Cases & Size $\ell$ & i-Mesh & $r_0$ & $\omega_{\rm in}$ & $k_f$ & $r_e$ & $\omega_0$ & $Z_0$ 
& $E_0$ &$\nu$ & f-mesh\\
P& $4\pi$   & $128^3$ &0.33& 0.6   & 8.4  &0.56 & 0.5 & $2.68$ & 0.67 & 2.5e-4:1.25e-4 & $512^3$  \\
P& $4\pi$   & $128^3$ &0.33& 0.6   & 8.4  &0.56 & 0.5 & $2.68$ & 0.67 & 6.25e-5:3.125e-5 & $1024^3$  \\
P& $6\pi$   & $256^3$ &0.33& 0.6   & 8.4  &0.56 & 0.5 & $2.68$ & 0.72 & 1.56e-5:7.8e-6 & $2048^3$\\
Q& $3\pi$   & $128^3$ &0.25& 1.26  & 11.9 &0.40 & 1 & $5.48$  & 0.96 & 5e-4 & $512^3$\\
Q& $3\pi$   & $128^3$ &0.25& 1.26  & 11.9 &0.40 & 1 & $5.48$  & 0.96 & 2.5e-4:6.25e-5 & $1024^3$\\
Q& $4\pi$   & $128^3$ &0.25& 1.26  & 11.9 &0.40 & 1 & $5.29$ & 0.85 & 5e-4:1.25e-4 & $1024^3$ \\
Q& $4\pi$   & $128^3$ &0.25& 1.26  & 11.9 &0.40 & 1 & $5.29$ & 0.85 & 6.25e-5 & $2048^3$ \\
Q& $6\pi$   & $256^3$ &0.25& 1.26  & 11.9 &0.40 & 1 & $5.30$ & 0.90 & 3.125e-5 & $2048^3$ \\
Q& $9-12\pi$& $256^3$ &0.25& 1.26  & 11.9 &0.40 & 1 & $5.30$ & 0.95-0.99 & 1.5625e-5:2e-6 & $2048^3$ \\
R& $4\pi$   & $128^3$ &0.175 & 2.5 & 16.8 &0.29 & 1.9 & $10$ & 0.97 & 2.5e-4:6.25e-5 & $2048^3$ \\
S& $4\pi$   & $128^3$ &0.125& 5  & 23.8 &0.20 & 4 & $17.7$ & 1.03 & 5e-4:6.25e-5 & $2048^3$ \\
S& $6-12\pi$& $256^3$ &0.125& 5  & 23.8 &0.20 & 4 & $17.8$ & 1.13-1.17 & 3.125e-5:7.8e-6 & $2048^3$ \\
v221 & $(4\pi)^3/2$ & $256^3$pts & 0.75 & 1.43 & 3.1 & 1.26 & 1 & 25.8 & 19.5 & 2e-3
& $512^3$ \\
v221 & $(4\pi)^3/2$ & $256^3$pts & 0.75 & 1.43 & 3.1 & 1.26 & 1 & 25.8 & 19.5 & 1e-3:5e-4
& $2\!\times\!1024^3$ \\ 
v222 & $(4\pi)^3$ & $256^3$pts & 0.75 & 1.43 & 3.1 & 1.26 & 1 & 27.1 & 21.2 & 2.5e-4
& $4\!\times\!1024^3$ \\ 
\end{tabular}
\caption{Parameters for initial conditions. i-Mesh: Initialisation mesh. 
$r_0$: Pre-filter radii in profile function \eqref{eq:Rosenhead}.
$\omega_{\rm in}$: Pre-filter initial peak vorticity.
$k_f$: Filter wavenumber in \eqref{eq:kfilter}.
$r_e$: Empirical post-filter radii of filaments \eqref{eq:kfilter}.
$\omega_0$: Post-filter Initial $\|\omega\|_\infty$ \eqref{eq:ominfty}.
$Z_0$: Initial enstrophy \eqref{eq:enstrophy}.
$E_0$: Initial kinetic energy \eqref{eq:energy}.
$\nu$: Viscosities. f-Mesh: Final mesh.
The initial helicity for all of the calculations is ${\cal H}(t=0)=7.67\times10^{-4}$ 
\eqref{eq:Hlink} using $\Gamma=0.505$. 
For anti-parallel cases (v221-v222), 
the symmetrised compuational domain sizes $L_x\!\times\!L_y\!\times\!L_z$ are listed. 
The periodic domain sizes would be $L_x\!\times\!2L_y\!\times\!2L_z$. 
The v221 $\nu$=2e-3 to 5e-4 cases use meshes of up to 
$1024\times1024\times2048$ points in $4\pi\times4\pi\times2\pi$ domains. 
The v222 $\nu=$2.5e-4 case uses a $2048^3$ mesh in a $(4\pi)^3$ domain.
The initialisation meshes all used $128\times256\times512=256^3$ points. 
}
  \label{tab:cases}
  \end{center}
\end{table}

The goal of the initialisation is to replicate the behaviour of the experimental trefoil knots
\citep{KlecknerIrvine2013,ScheeleretalIrvine2014a}, all of which have a single
dominant initial reconnection. 
To do this one needs to weave a single vortex of finite diameter and fixed circulation into 
a perturbed (2,3) knot, a knot with a self-linking of ${\cal L}_S=3$ \eqref{eq:selflink}
due to three crossings over two loops, as shown in figure \ref{fig:T6}. To compare to the experiments,
it should not generate three simultaneous weak reconnections, 
as found by the symmetric trefoil calculations of \cite{KidaTakaoka87}. Instead, it 
should be perturbed so that it reproduces the experimental dynamics with a single major
reconnection event to allow comparisons between the simulated continuum global helicity 
\eqref{eq:helicity} and the experimental centre-line helicity diagnostics. 

The trefoil trajectory in this paper is defined by:
\EQL{eq:trefoil}\begin{array}{rrl} x(\phi)= r(\phi)\cos(\alpha) &\quad
y(\phi)= & r(\phi)\sin(\alpha) \qquad z(\phi)= a\cos(\alpha) \\
{\rm where} & r(\phi) =& r_f+r_1a\cos(\phi) +a\sin(w\phi+\phi_0)\\
{\rm and} &  \alpha=& \phi+a\cos(w\phi+\phi_0)/(wr_f) \end{array} \EN
with $r_f=2$, $a=0.5$, $w=1.5$, $\phi_0=0$, $r_1=0.25$ and $\phi=[1:4\pi]$.  This 
weave winds itself twice about the central deformed ring with: $x_c^2(\phi)+y_c^2(\phi)=r_c^2(\phi)$,
where $r_c(\phi)=r_f+r_1a\cos(\phi)$  for a $r_1\neq0$ perturbation. 
The separation through the $r=r_c$ ring of the two loops of the trefoil is $\delta_a=2a=1$. 
Four additional low intensity vortex rings, two moving up in $z$ and two down, 
provided the perturbation that breaks the three-fold symmetry of the trefoil 
so that it has a single major initial reconnection like the experiments.  

While the $r_1\neq0$ term was added with the intention of generating only one
reconnection, in all cases the perturbation disappeared before 
any reconnections began.  It was then pointed 
out\footnote{A.W. Baggaley and C.F Barenghi, private communication, 2015.} 
that the platform that the trefoil model was placed upon probably generated additional 
independent vortices that would not have been identified by the hydrogen bubbles shed by
the primary 3D-printed knot. 
Therefore, the four low intensity vortex rings in table \ref{tab:extrarings},
each propagating either in $+z$ or $-z$ were placed on the periphery of the trefoil to give it 
the type of external perturbation that the platform might have generated.

{\bf Profile and direction} 
After the trajectories of the trefoil and the four auxilary rings are established, the vorticity 
vectors are mapped onto the computational mesh using a modification of the method described 
in \cite{Kerr2013a}. This starts by identifying the closest locations on each filament 
for every mesh point and the distances between these filament locations and the mesh points. 
This is used exactly for the auxilary rings.

A modification is needed for initialising a trefoil because two points on the trefoil's core 
trajectory have to be identified for every point on the computational mesh, 
One point from each of the trefoil's two loops, one point from $\phi=[0:2\pi]$ and
one point from $\phi=[2\pi:4\pi]$ in \eqref{eq:trefoil}.  To avoid overcounting, the space 
perpendicular to the central $z$ axis is divided into octants. First, the octant
containing the mesh point is found. Next, to find the nearest points on the two loops about
the trefoil's core to this mesh point, one restricts the search to those points on the loops 
that pass through that octant and the two octants on either side.

\begin{table}
\vspace{-3mm}
\bminic{0.25}
  \begin{center}
\begin{tabular}{cccc}
$\omega_{\rm in}$ & $x$ & $y$ & $z$ \\
0.075 & -0.5 & 0.25 & 0.75 \\
-0.075 & 0.5 & 0.5 & -0.75 \\
-0.075 & -0.5 & -0.45 & 0.95 \\
0.075 & 0.8 & -0.5 & 0.95 
\end{tabular} \end{center}\emini~~~~\bminic{0.73}
\caption{Every trefoil used the same set of four additional rings placed around the trefoil 
to generate the pertubations that yielded a single, dominant initial reconnection. For each one,
the ring radii were $r_f=2$, the core radii were $r_0=0.5$ and the pre-filtering 
$|\omega_{\rm in}|=0.075$. The table gives the signs of the $\omega_{\rm in}$ 
and their positions with respect to the centre of the trefoil.}
\emini\\ ~ \\ \vspace{-3mm}
  \label{tab:extrarings}
\end{table}

Once the nearest points on the trefoil loops to a mesh point have been identified, 
the mapping of the trefoil vortex's direction and profile function closely follows the method in
\cite{Kerr2013a}. 
The profile function used for the pre-filter vorticity $\bomega_i(r)$ is based upon 
the Rosenhead regularisation of a point vortex: 
\EQL{eq:Rosenhead} |\bomega_i|(r)=\omega_{\rm in}\frac{16r_0^4}{(r^2+4r_0^2)^2}\,. \EN
This is followed by smoothing the resulting vector field $\bomega_i(t)$ using a hyperviscous 
Fourier filter
to get the final transformed  vorticity field $\bomega_f(k)$ using
\EQL{eq:kfilter} \bomega_f(k)=\bomega_i(k)\exp(-k^4/k_f^4)\quad
\mbox{with resulting in filtered radii}~
r_e=\bigl(\Gamma/(\omega_0/\pi)\bigr)^{1/2} \,. \EN
The smoothing radii $r_0$ and filter wavenumbers $k_f$ are given in table \ref{tab:cases}. 
The parameters $\omega_{\rm in}$, $r_0$ and $k_f$ were chosen such that the circulation of all 
the initial trefoil vortices is $\Gamma=0.505$ and the effective radii of the vortices
are $r_e\approx1.6 r_0$. 

This initialisation was done on $128^3$ meshes for the $(3\pi)^3$ and $(4\pi)^3$ domains and 
on $256^3$ meshes for the larger domains. To get to the computational meshes, 
the $128^3$ and $256^3$ data was remeshed by filling higher-wavenumber modes with zeros.
Figure \ref{fig:T6} at $t=6$ shows a Q-trefoil (see table \ref{tab:cases}) shortly after the 
calculation began.  The two diagnostics are an isosurface of 0.55$\|\omega\|_\infty$ and 
two closed filaments \eqref{eq:vortexlines} seeded at or near the position of 
$\|\omega\|_\infty$ in the centre of the trefoil. These curves  were then used in \eqref{eq:Gauss} to
verify that the self-linking number \eqref{eq:selflink} is ${\cal L}_S=3$.

One of the goals achieved by this mapping is that for this initial vortex state 
there are almost no changes ($\leq0.5\%$) to the vorticity norms as the domain is
increased.  This is true for the helicity ${\cal H}$, cubic velocity norm $\sbl{u}{L^3}{}$ 
\eqref{eq:LpHsT3R3}, the initial enstrophy $Z$ and the peak vorticity $\sbl{\omega}{\infty}{}$ 
once the domain was at least $(4\pi)^3$.  
The only property that changes appreciably as the domain is increased is the kinetic energy, 
which increased by 5\% between each of the
$(4\pi)^3$, $(6\pi)^3$, $(9\pi)^3$ and $(12\pi)^3$ domains. 
Because multiple properties of the initial state are independent of the size of the domain,
when run in different domains, the results were nearly identical so long as the periodic
boundaries did not interfere with the interactions.

\subsection{Resolution and viscosity range\label{sec:viscousrange}} 

Once a perturbation was identified that qualitatively reproduced the experimental reconnection 
with a strong, single, initial reconnection, as discussed in section 
\ref{sec:helicity}, it was realised that the growth of $\sqrt{\nu}Z(t)$ 
up to a common time of $t_x\approx40$ was
independent of $\nu$ for $\nu\geq1.25$e-4. To go beyond that modest range of viscosities
($\nu=$5e-4 to 1.25e-4), it was then discovered that by increasing the domain size $\ell$ 
as $\nu$ was decreased further, the crossing at $t_x\approx40$ could be continued to even 
smaller $\nu$. This effect places the two opposing demands upon the available resolution 
at small and large scales that needs to be discussed before the resulting self-similar collapse 
is presented in section \ref{sec:firstrecon}

The two opposing demands stem from \cite{Constantin86}, which provides the conditions under 
which small $\nu$ Navier-Stokes solutions will converge to the Euler solutions. 
As discussed briefly in section \ref{sec:increasingell},
these conditions apply when $\nu<\nu_s(\ell)$ \eqref{eq:nus}, where the domain dependent
critical viscosities $\nu_s(\ell)\to0$ as $\ell\to\infty$, This dependence of $\nu_s$ upon $\ell$ 
means that one can relax any upper bounds on growth set by the $\nu_s$ by increasing $\ell$. 
Figure \ref{fig:QsnuZ}a, demonstrates both effects. First, how the $\nu_s(\ell)$ constraint 
for fixed $\ell$ can supress the scaling of $\sqrt{\nu}Z(t_x)$ as $\nu$ is decreased. Then how
it can be relaxed by increasing $\ell$. This is discussed further in section \ref{sec:firstrecon}.

\begin{figure} 
\includegraphics[scale=0.32]{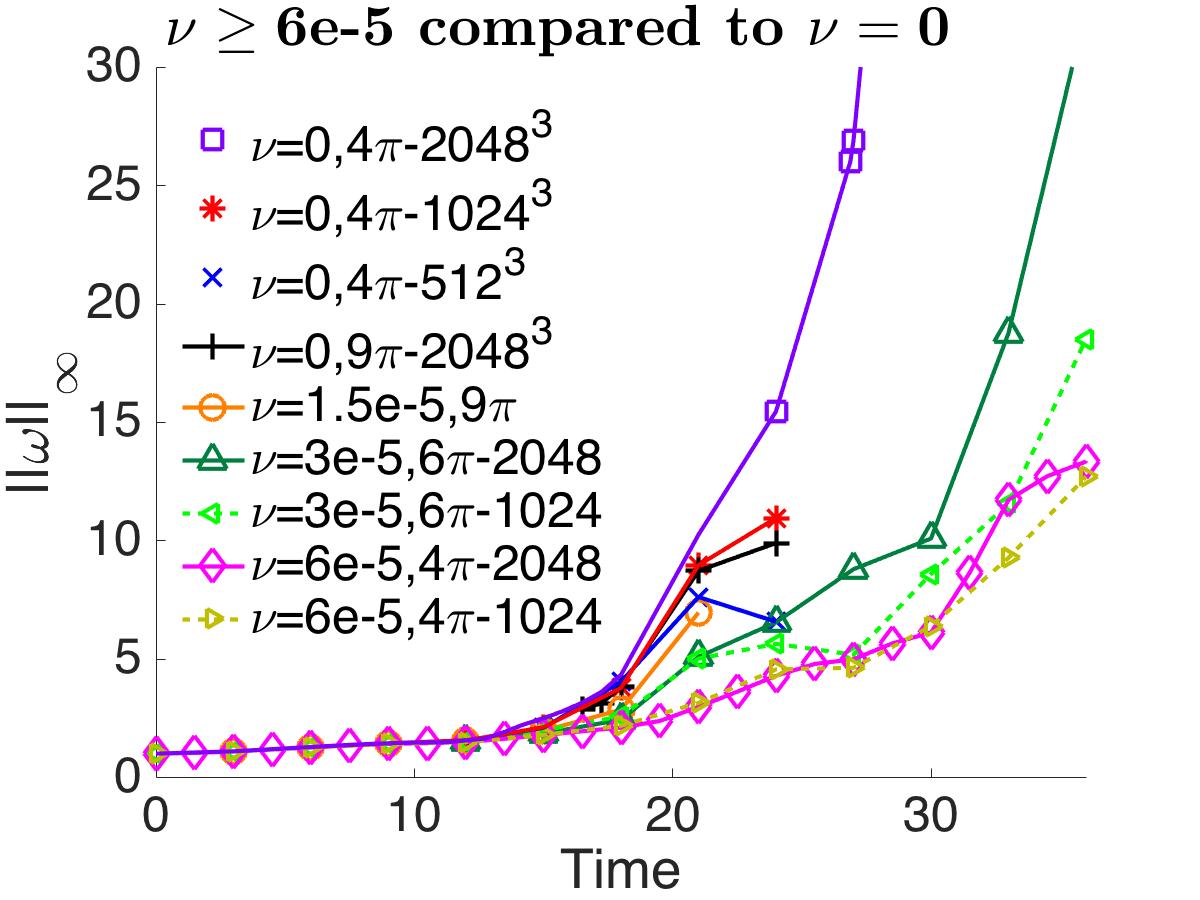}\\
\caption{\label{fig:EulerNSomaxtsmall} Early time $\|\omega\|_\infty$ for Euler and 
small $\nu$ Navier-Stokes calculations. The Navier-Stokes $\|\omega\|_\infty$ are used for 
resolution checks in section \ref{sec:viscousrange} and all of the $\|\omega\|_\infty$ 
curves are used in section \ref{sec:regularity} to address possible mathematical constraints 
upon the new scaling behaviour. There are four Navier-Stokes cases using two
viscosities, $\nu=6.25$e-5 with $\ell=4\pi$ and $\nu=3.125$e-5 with $\ell=6\pi$, 
with each calculated on both $1024^3$ and $2048^3$ meshes.
There are three Euler calculations in a $\ell=4\pi$ domain at resolutions of 
$512^3$, $1024^3$ and $2048^3$ and one $\ell=9\pi$, $2048^3$ Euler calculation. 
}
\end{figure}
\begin{figure} 
\includegraphics[scale=0.32]{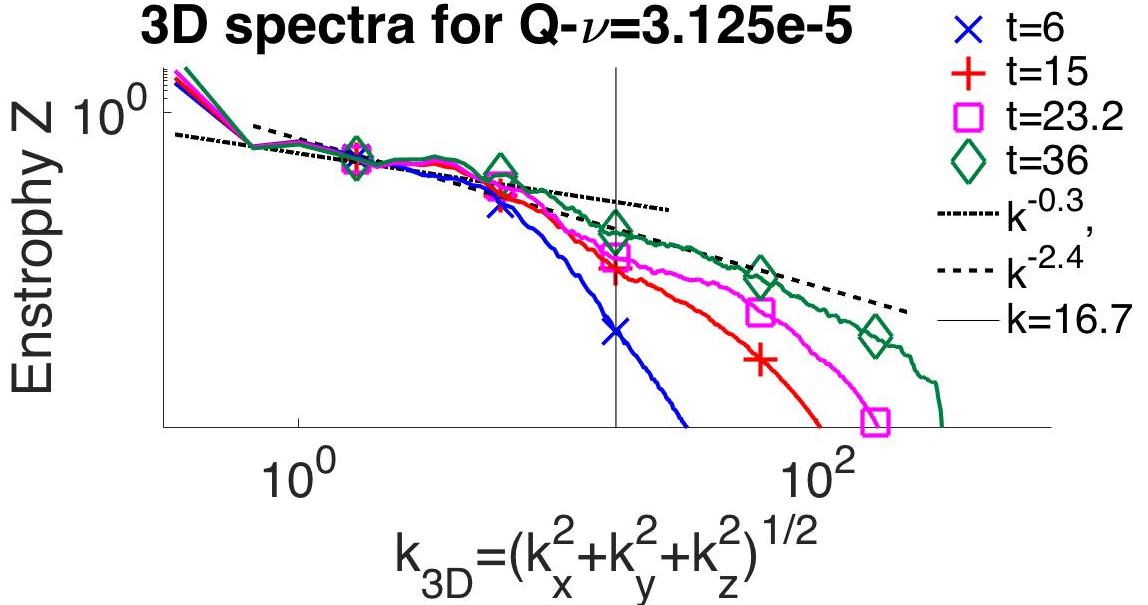}
\caption{\label{fig:NSomspectra} 
For provide resolution checks, Navier-Stokes enstrophy spectra for the
$\nu=3.125$e-5, $2048^3$ calculation at $t=6$, 15, 23.2 and 36.
There are two spectral regimes. A growing, nearly flat lower-wavenumber $k^{-0.3}$ regime
and higher-wavenumber $Z(k)\sim k^{-2.4}$ (or energy $E(k)\sim k^{-4.4}$). 
}
\end{figure}

Two approaches have been used to determine the limits 
in time and decreasing viscosity over which the diagnostics can be believed for
a given resolution and domain size. One approach, shown in figure \ref{fig:EulerNSomaxtsmall}, 
is to compare the evolution of $\|\omega\|_\infty$ for different resolutions and viscosities 
The other approach is to follow the evolution of enstrophy spectra 
such as those from the $\nu=3.125$e-5 case in figure \ref{fig:NSomspectra}. 
The Navier-Stokes comparisons will be discussed next. The discussion of the Euler 
calculations and the comparisons between Euler and Navier-Stokes are
in section \ref{sec:regularity}.

For Navier-Stokes comparisons, figure \ref{fig:EulerNSomaxtsmall} uses two viscosities, 
$\nu=3.125$e-5 and $\nu=6.25$e-5 with two resolutions for each, $1024^3$ and $2048^3$.  
Of these four calculations, even though the $\nu=6.25$e-5, $\ell=4\pi$ $2048^3$ calculation is
the only one judged to be fully resolved in terms of $\|\omega\|_\infty$ and spectra, 
all except the $\nu=3.125$e-5, $\ell=6\pi$, $1024^3$ case provide reliable 
results for the enstrophy $Z$.

The $\nu=3.125$e-5 three-dimensional enstrophy spectra shown in figure \ref{fig:NSomspectra} 
can be used to assess whether calculations are adequately resolved as follows. First note
that for this $\nu=3.125$e-5 case there are exponential tails to the spectra up to $t=24$, 
a signature that they are fully resolved. 
For the $\nu=6.25$e-5 $\ell=4\pi$ cases (not shown), the $2048^3$ spectra
have exponential tails for all times, but for the $1024^3$ case, there are spectral tails 
only up to $t=24$, similar to those in figure \ref{fig:NSomspectra}. Due to this,
between $t=31$ and $t=36$, the $\|\omega\|_\infty$ for the $1024^3$ case are below those for
the $2048^3$ case in figure \ref{fig:EulerNSomaxtsmall}.  Furthermore, at no time are there any 
discernible differences in the enstrophies, $Z$ for the $\nu=6.25$e-5, $1024^3$ and $2048^3$ cases.

Why doesn't the drop in $\|\omega\|_\infty$ due to inadequate resolution affect the growth
of the enstrophy $Z$ for the $\nu=6.25$e-5, $1024^3$ and $3.125$e-5, $2048^3$ calculations?
This can be understood by noting how the growing enstrophy-containing, lower-wavenumber, 
approximately $Z(k)\sim k^{-0.3}$ regime gradually expands into 
the $Z(k)\sim k^{-2.4}$ power law regime.  A regime whose magnitude increases, 
but slope does not, as it extends to the highest wavenumbers, 
as indicated for $t=36$. This $k^{-2.4}$ regime shields the $k^{-0.3}$ enstrophy-containing
wavenumber regime from the high-wavenumber cutoff and small-scale errors. This effect also
applies to the early-time ($t\leq24$) very small $\nu$
enstrophies plotted in figure \ref{fig:EulerNS-Zttx} and discussed in section \ref{sec:regularity}.
\RMKA{If the $k^{-2.4}$ regime were exactly $k^{-2}$ would be the least steep that can 
give finite helicity dissipation.  The $\nu=1.5125e-6$ case has adequate resolution 
for $t\leq21$ and was selected to show that $\|\omega\|_\infty$ is bounded the Euler 
$\|\omega\|_\infty$ for all the smaller $\nu$ calculations down to $\nu=2e-7$.}

Based upon these observations, because the $\nu=3.125$e-5, $\ell=6\pi$, $2048^3$ resolution case 
is mildly underresolved for a brief period after $t\approx30$, the $\nu<3.125$e-5 cases
are not among the primary rescaled enstrophy curves in figures \ref{fig:QsnuZ} 
and \ref{fig:QHdRisnuZtime}.  
The exception in each figure is one gray, dashed $\nu=2$e-6 curve, that shows how robust 
the small $\nu$ scaling of $\sqrt{\nu}Z(t)$ is for small $\nu$ up to $t=t_x\approx40$
and connects the early time,
very small viscosity cases given in figure \ref{fig:EulerNS-Zttx} to the higher viscosity cases.

\subsection{Trefoil length and time scales \label{sec:LTscales}}

To compare these simulations to the experiments in section \ref{sec:helicity}.and 
earlier simulations, the important length and time scales are needed. 
The three length scales in the initial condition \eqref{eq:trefoil} are the trefoil's radius $r_f=2$, 
the separation between its loops $\delta_a=2a=1$ and the effective
thickness of the filaments $r_e$.  Four $r_e$ are simulated, designated P, Q, R and S, 
each halving the area of the trefoil core of the previous set of initial conditions while
keeping the circulation $\Gamma$ constant.  The focus will be on the Q-trefoils, 
with results from the S-trefoils used to demonstrate that these results do not depend 
strongly upon the initial core radii $r_e$ for $r_e/r_f$ that 
are close to those used by the experiments discussed in section \ref{sec:helicity}.
\DEL{(relative $r_e$ is about twice the
experimental value)\footnote{Irvine private communication, August 2016}.
and to show that the enstrophy at early times and low viscosities is not bounded by the Euler enstrophy.}
\DEL{For the Q and S initial conditions, the smallest properly-resolved viscosity, 
$\nu=3.125\times10^{-5}$ gives a Reynolds number of $Re_\Gamma=\Gamma/\nu=16095=1.6e4$.  For 
the under-resolved case with the smallest viscosity $\nu=2\times10^{-6}$ the Reynolds number is $Re=2.5e5$.}
\DEL{and P initial states respectively to: Demonstrate}

There are a variety of timescales that can be applied to vortex reconnection events, either
nonlinear, viscous or maybe a combination of the two. The results for scaled enstrophy 
$\sqrt{\nu}Z$ growth in figure \ref{fig:QsnuZ} and helicity evolution in figure \ref{fig:HL3H12}
will show that the correct dynamical timescale should
not depend upon either the domain size $\ell$ or the core thickness $r_e$. 
A traditional large-eddy turnover time $t_L$ can be determined using the mean velocity scale 
of the energy and the size of the structure $r_f$
\EQL{eq:te} t_L=r_f/U,~{\rm where}\quad U=\sqrt{2E_0}\,.\EN
Table \ref{tab:cases} shows that $E_0$ is a function of domain $\ell$ and the
thickness of the cores $r_e$, so $t_L$ depends upon $\ell$, $r_e$ and $r_f$ and
is not an appropriate nonlinear timescale.

A better timescale for these simulations
is to use the strength of the nonlinear convective motion, given by the circulation 
$\Gamma$ and the size of the structure. Using $\Gamma$ and
either $r_f$ or $a$ gives the following two nonlinear timescales for the trefoils:
\EQL{eq:ta} t_f=r_f^2/\Gamma=8\quad{\rm and}\quad   t_a=\delta_a^2/\Gamma=t_f/4=2     \,.\EN 
$t_f$ will be used for comparisons with the experiments and $t_a$, based upon the separation of
the loops, will be used for comparisons between the trefoil and anti-parallel reconnection
in section \ref{sec:antip}.

Furthermore, using $\Gamma$ and $r_f$ gives the characteristic velocity scale of
\EQL{eq:uf} u_f=\Gamma/r_f=0.505/2=0.25\,. \EN
The initial peak velocity and downward ($-z$) translation velocity of the position of
$\|\omega\|_\infty$ are $\|u\|_\infty\approx u_z(\|\omega\|_\infty)\approx -0.25=-u_f$.
The velocities $\|u\|_\infty$ and $|u_z(\|\omega\|_\infty)$ decrease slightly until $t=10$, then 
increase. The change in peak velocity behaviour at $t=10$ could be a precursor to the
$(\sqrt{\nu}Z(t))^{-1/2}$ collapse for the trefoils discussed in section \ref{sec:SS}.

However, if the trefoils' nonlinear timescale is  $t_f=8$,  the most important physical time 
is $t_x\approx40=5 t_f=20t_a$, which is when the $\sqrt{\nu}Z(t)$ approximately meet in 
figure \ref{fig:QsnuZ} and when the first reconnection ends, as discussed in section 
\ref{sec:reconnection}.  Resolving these incongruous timescales is one of the goals of this paper.


\section{Scaling during first reconnection at $t_x$\label{sec:firstrecon}}

The first indications for a new type of $\sqrt{\nu}Z(t)$ scaling for trefoil reconnection 
came from the $\nu$-independent crossing of $\sqrt{\nu}Z(t)$ at $t=t_x\approx40$ for the first 
five Q-trefoils in figure \ref{fig:QsnuZ}a.
The two vertical lines show $t_x$, the time for the $\sqrt{\nu}Z(t)$ convergence and 
$t_\epsilon\approx 2t_x$, a later, more approximate time when a common value for
the energy dissipation rates $\epsilon=\nu Z$ is attained.  Frame \ref{fig:QsnuZ}b 
highlights this second convergence by plotting $\epsilon$ for three viscosities.

The first four cases using a $(4\pi)^3$ domain cover $\nu=5$e-4 to $\nu=6.25$e-5 while 
the green $\nu=3.125$e-5 calculation uses a $(6\pi)^3$ domain.  There are two additional 
$\sqrt{\nu}Z(t)$ evolution curves in the main frame.  The gray-dashed $\nu=$2e-6 curve and
the brown-marked $\nu=$3.125e-5 curve that comes from a $\nu=3.125$e-5 calculation using a smaller 
$\ell^3=(4\pi)^3$, domain than the $(6\pi)^3$ calculation whose $\sqrt{\nu}Z(t)$ crosses
with the other $\nu<$3.125e-5 curves at $t_x=40$.

This dependence of $\sqrt{\nu}Z(t)$ on $\ell$ is general. That is, when 
$\nu=$6.25e-5 and a $(3\pi)^3$ domain was used, $\sqrt{\nu}Z(t_x)<0.14$, but with
the $(4\pi)^3$ domain it does cross the others in figure \ref{fig:QsnuZ}a
with $\sqrt{\nu}Z(t_x)=0.14$.
As the viscosity is decreased further, to $\nu\leq1.5625$e-5, the $\sqrt{\nu}Z(t_x)$ converge 
at $t_x\approx40$
only if $\ell$ is increased still further. First to $\ell=9\pi$ and then to $\ell=12\pi$, 
as in the $\nu=2$e-6 gray-dashed curve that is included to demonstrate the robustness of the 
of the convergence of $\sqrt{\nu}Z(t)$ at 
$t=t_x\approx t_x\approx40$ even when the viscosity has changed by a factor of at least 256 
and the calculation is clearly under-resolved in terms of $\|\omega\|_\infty$. 

\subsection{Mathematics underlying increasing $\ell$ \label{sec:increasingell}}

Why must the domain be increased as the viscosity is decreased? A plausible answer comes
from considering this question:
As $\nu$ decreases, could there be critical viscosities $\nu_s$ such that for
$\nu<\nu_s$, the Navier-Stokes norms are bounded? 

This question was considered by \cite{Constantin86} who showed that if in a fixed periodic domain 
the Euler solution for a particular initial condition is regular (non-singular) 
up to a time $T$, then for $\nu<\nu_s$, a critical viscosity, the higher-order $s>5/2$ 
Sobolev norms \eqref{eq:HsV} for the differences between the Navier-Stokes $u(t)$ 
and Euler $v(t)$ solutions are bounded as follows:
\EQL{eq:Hsbnd} \sup_{t\in[0,T]}\sbl{u(t)-v(t)}{H^s_\ell}{}\leqq \nu_s\gamma_s\,, \EN
where the $\gamma_s$ are viscosity independent functions of the regular Euler norms and 
the $\nu_s$ depend upon $\ell$ as follows 
\EQL{eq:nus} \nu_s(\ell)\sim \ell^{-2s+5}\left\{\mbox{functions of Euler time integral}\right\}
\rightarrow 0 ~~{\rm as}~~\ell\rightarrow\infty~~{\rm for}~~s>5/2\,. \EN
Once these bounds on the higher-order $\sbl{u}{\dot{H}^s_\ell}{}$ norms are established, 
then a chain of $\ell$ dependent Sobolev space embedding inequalities can show that these norms
would in turn bound $\sbl{\omega}{L^\infty}{}$ from above. That the early time Navier-Stokes
$\sbl{\omega}{L^\infty}{}$ are bounded at early times by the non-singular Euler
$\sbl{\omega}{L^\infty}{}$ is shown in figure \ref{fig:EulerNSomaxtsmall}. 
From this, the standard H\"older inequality will bound the volume-integrated enstrophy 
$Z\leq \ell^3\sbl{\omega}{\infty}{2}$ \eqref{eq:Holder}, with a very strong $\ell$-dependent
pre-factor. \hfill $\square$

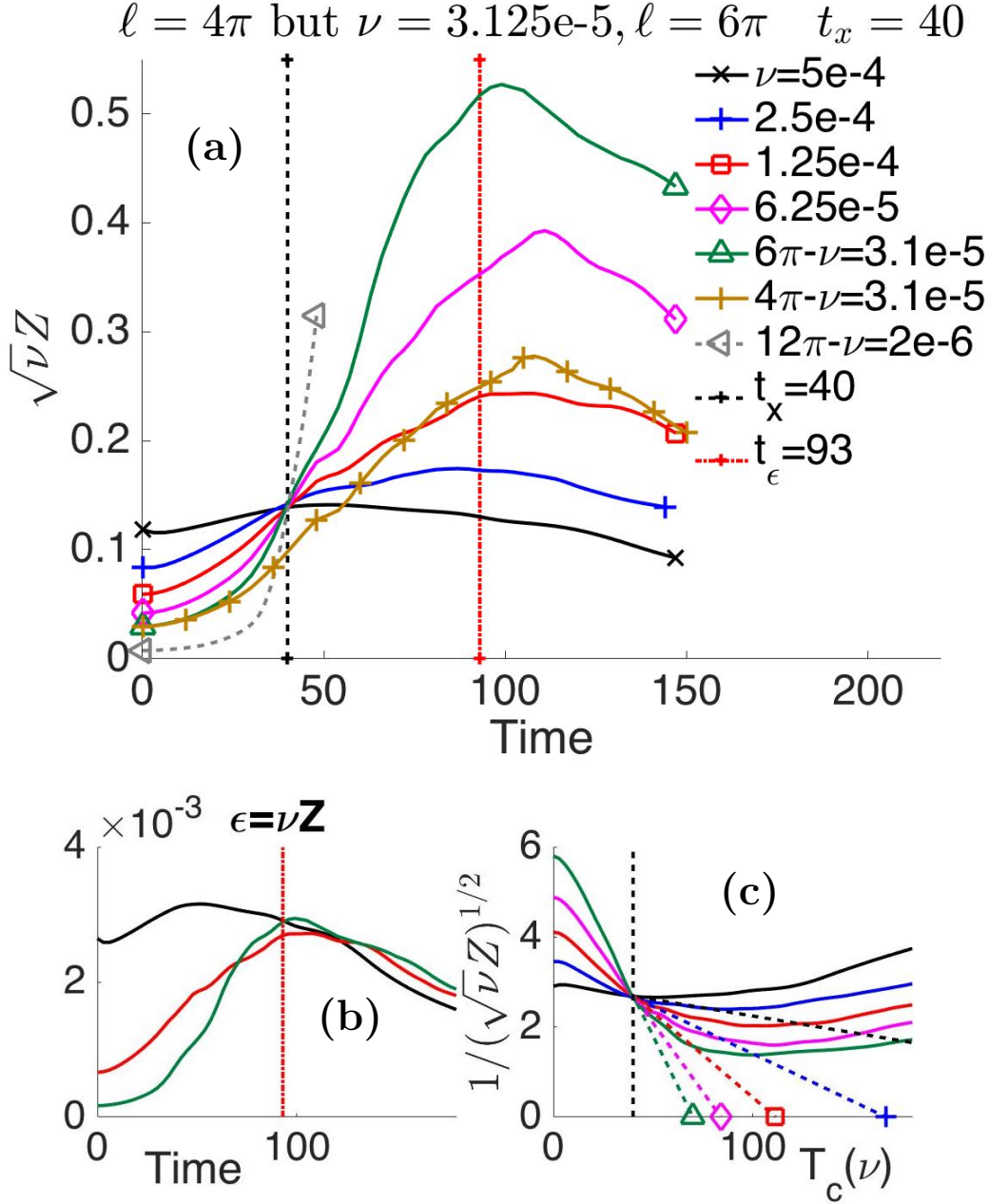
\begin{figure}
\includegraphics[scale=0.34]{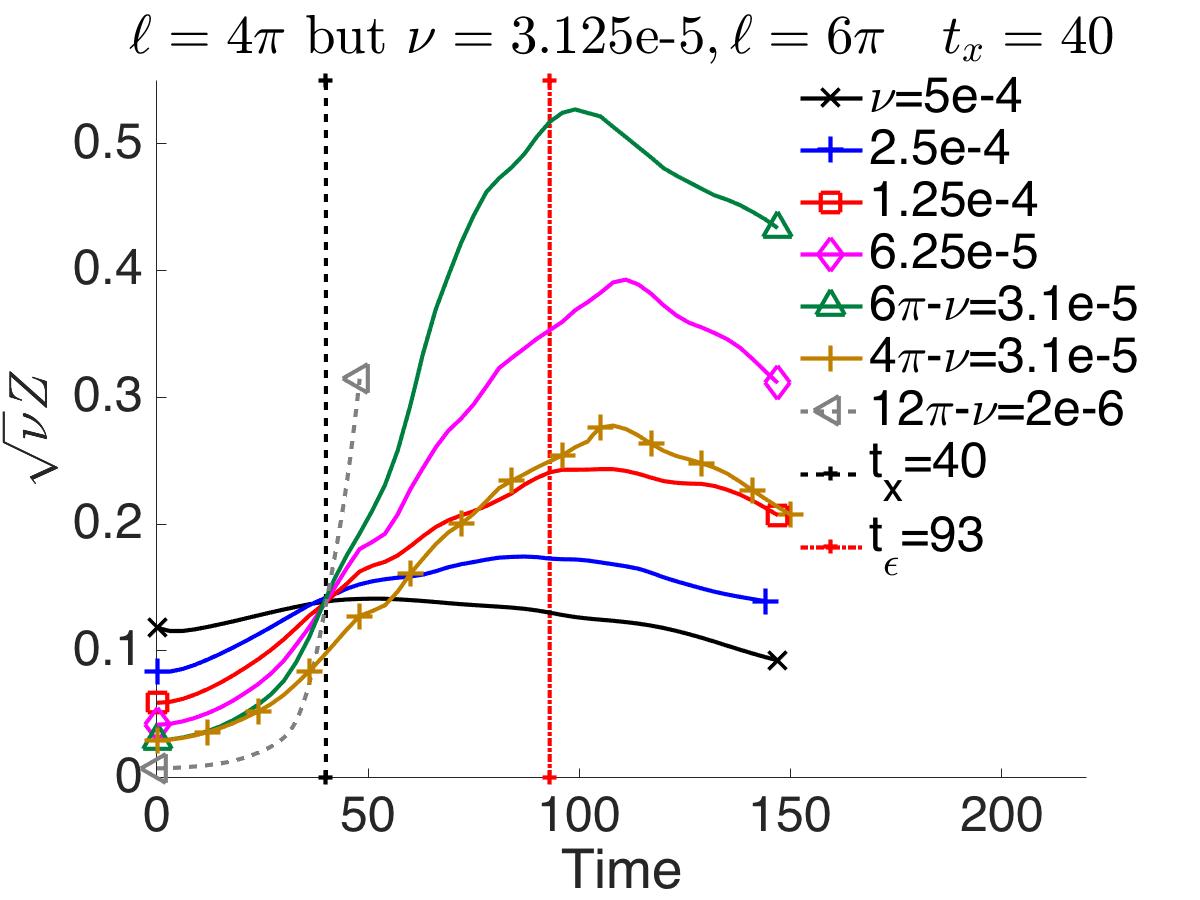}
\begin{picture}(0,0) \put(70,250){{\LARGE\bf (a)}}\end{picture} \\
\bminic{0.05}~\emini\bminic{0.95}
\includegraphics[scale=0.32]{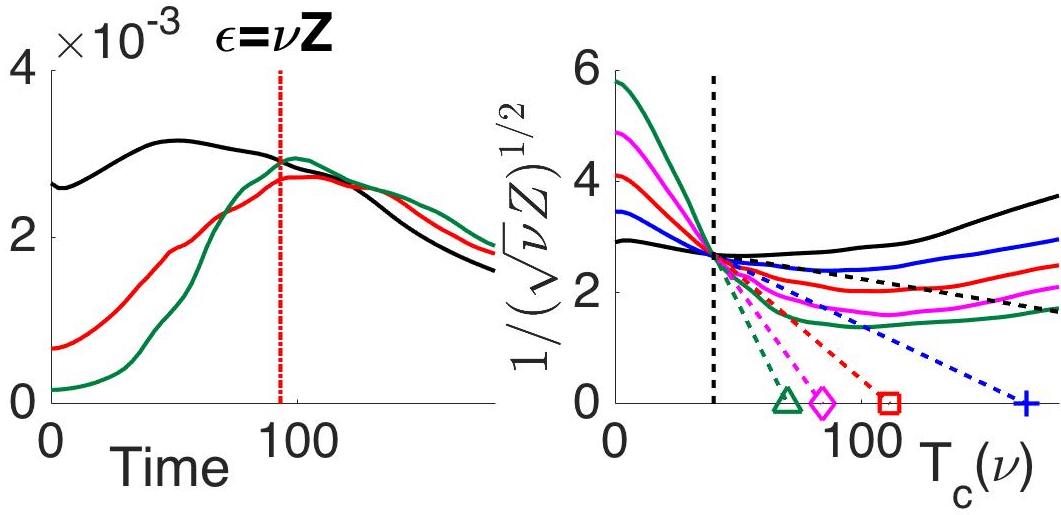}
\begin{picture}(0,0)\put(-240,70){{\LARGE\bf (b)}} 
\put(-80,120){{\LARGE\bf (c)}}\end{picture} 
\emini
\caption{\label{fig:QsnuZ} 
{\bf a:} Time evolution of the scaled enstrophy $\sqrt{\nu}Z$.  
Note that all the curves, except the brown-+ curve, cross at 
$\sqrt{\nu}Z(t_x)=0.14$, $t_x\approx40$. The time $t_x=40$ is identified as the end of the 
first reconnection using graphics that includes figure \ref{fig:T45} at $t=45$.  
The brown-+ curve is from a $\nu=3.125$e-5, $(4\pi)^3$ calculation, the
same domain as the other viscous cases, but $\sqrt{\nu}Z(t=40)\neq 0.14$.
Unlike the $\nu=3.125$e-5 green curve which was run in a $(6\pi)^3$ domain.
The importance of increasing the size of the domain as the viscosity is decreased comes
from the mathematical analysis of \cite{Constantin86} expressed by \eqref{eq:nus}. 
The red line at $t_\epsilon=93$ is when plots of 
the dissipation rate $\epsilon=\nu Z$ cross.
{\bf b:} $\epsilon=\nu Z$ for three cases, $\nu=5$e-4, 1.25e-4 and 3.125e-5
that show it crossing at $t_\epsilon=93$.
{\bf c:} This shows the transitional plot that determines $T_c(\nu)$ 
for the time scaling in figure \ref{fig:QHdRisnuZtime}. $T_c(\nu)$  is found by 
plotting $B_\nu(t)=(\sqrt{\nu}Z(t))^{-1/2}$ \eqref{eq:Bnu}, then drawing
lines between $t_\Gamma=15$ and $t_x$ that are extended to $B_\nu(t)=0$, giving us 
the effective singular times $T_c(\nu)$ indicated by the symbols in the upper caption.
The $\Delta t(\nu)=T_c(\nu)-t_x$ are used in the \eqref{eq:isnuZBxtime} scaling of $B_\nu(t)$ 
that is plotted in figure \ref{fig:QHdRisnuZtime}. 
}
\end{figure}

\begin{figure} 
\includegraphics[scale=.32]{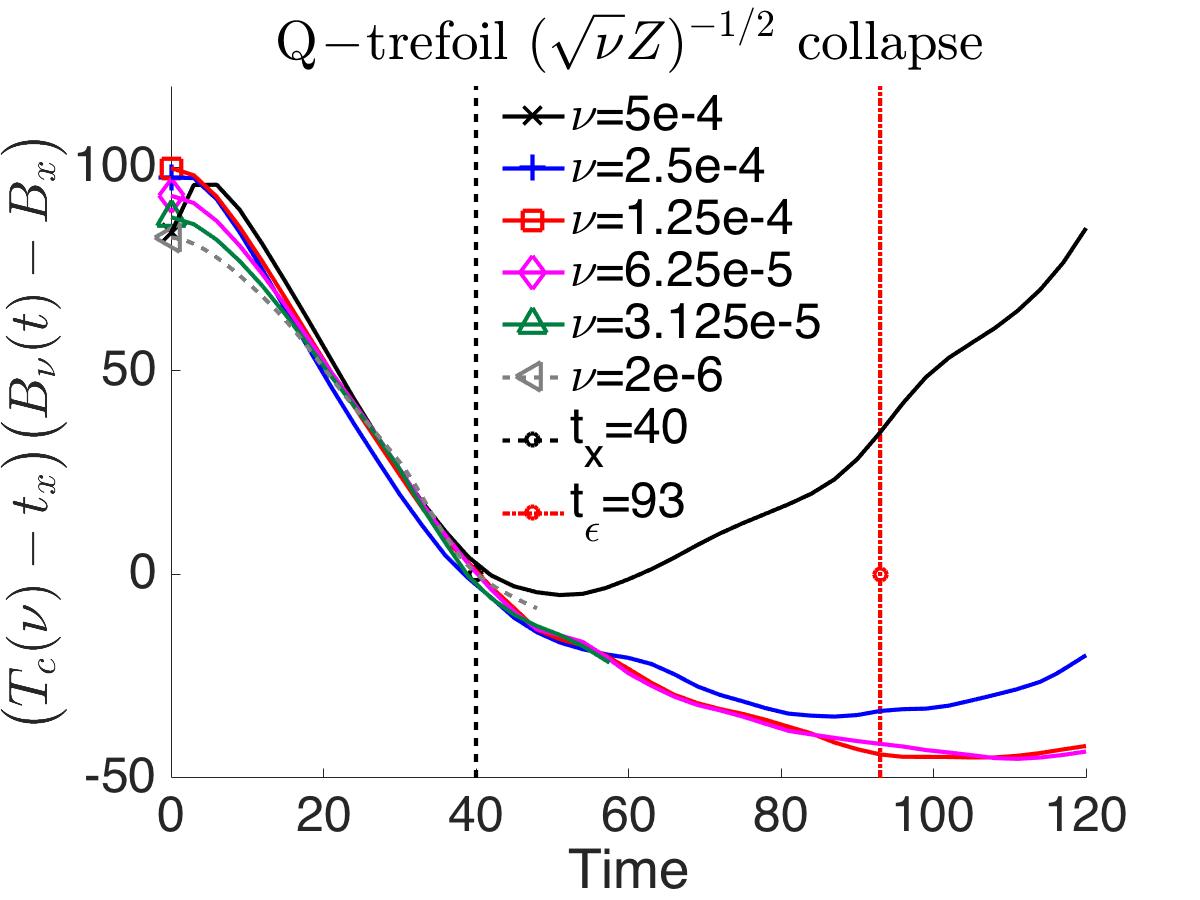}
\caption{\label{fig:QHdRisnuZtime} 
Evolution of the inverse, scaled enstrophy using \eqref{eq:isnuZBxtime}
with $B_\nu(t)=(\sqrt{\nu}Z(t))^{-1/2}$ \eqref{eq:TcDt} and $B_x=B_\nu(t_x)$.
The left symbols are $t=0$ for each $\nu$ calculation.
This collapse to early times indicates that the dynamics leading to the crossing of 
$\sqrt{\nu}Z(t)$ at $t=t_x$ begins relatively early and even extends to $t>t_x$.
The collapse begins at $t_\Gamma=15$, an early time relative to $t_x=40$, begins as soon as
negative helicity appears in the vicinity of the closest approach of the two loops
as discussed in section \ref{sec:Hnegative}.
That viscosity plays a role almost immediately as discussed in
section \ref{sec:regularity} using figure \ref{fig:EulerNS-Zttx}.  }
\end{figure}

$\bullet$ Note that despite the \cite{Constantin86} restriction on the growth of 
enstrophy $Z$ as $\nu\rightarrow0$ for fixed $\ell$,
because the $\nu_s(\ell)$ depend inversely upon the domain size $\ell$, 
those restrictions can be relaxed simply by making the domain larger. 
Figures \ref{fig:QsnuZ} and \ref{fig:QHdRisnuZtime} show how that allows $\sqrt{\nu}Z(t)$ 
to collapse onto a self-similar curve  as $\nu$ decreases.
Given this, the following questions are to be addressed in the following sections:
\ENM\item How will simulations with smaller viscosities, in larger domains,
behave? 
\ITM\item[$\circ$] Small viscosities will be considered in section \ref{sec:regularity}. \ITN
\item What type of physical space mechanism could connect the distant periodic 
boundaries to the enstrophy generation mechanisms within the original trefoil?
\ITM\item[$\circ$] A physical space mechanism will be considered with respect
to the negative helicity isosurface in figure \ref{fig:T45} at $t=45$ and in 
section \ref{sec:helicity}.  \ITN
\item Why does the growth of $\sqrt{\nu}Z$ have self-similar scaling instead of the dissipation 
$\epsilon=\nu Z$? 
\ITM\item[$\circ$] 
This and whether there could be any type of self-similar collapse in the period
$0\lesssim t\leq t_x$ are now discussed in section \ref{sec:SS}.\ITN
\EEN

\subsection{Self-similar scaling in time \label{sec:SS}}

After searching for, and not finding, any clear indications that the energy dissipation
rate $\epsilon=\nu Z$ might have a direct role in the reconnection process, 
the $\sqrt{\nu}Z(t)$ diagnostic was explored based upon the anti-parallel circulation 
exchange results illustrated in figure \ref{fig:GammasepsG} and discussed in 
section \ref{sec:antip}. Both the circulation exchange rate for anti-parallel reconnection
$\epsilon_\Gamma$ \eqref{eq:dGammadt} and the $\sqrt{\nu}Z(t)$ diagnostic have
the earmarks of \cite{Leray34} scaling, as discussed in section \ref{sec:Leray}.

While seeing the crossing of all the $\sqrt{\nu}Z(t)$ in figure \ref{fig:QsnuZ}a 
at a $\nu$-independent time $t_x$ is tantalising, the observation would be more significant 
if it could be associated with a self-similar collapse. The path to finding a
self-similar collapse for $t\leq t_x$ is in figure \ref{fig:QsnuZ}c which, in addition to showing
$t\leq t_x$ linearly decreasing 
\EQL{eq:Bnu} B_\nu(t)=\bigl(\sqrt{\nu}Z(t)\bigr)^{-1/2} 
=\bigl(\nu^{1/4}\sbl{\omega(t)}{L^2(\bbT_\ell^3)}{}\bigr)^{-1/2}\,,\EN
shows linear extrapolations of $B_\nu(t)$ from the earliest time that $B_\nu(t)$ is linear, 
designated $t_\Gamma\approx15$ for each Q-trefoil, through $t_x$ to the linearly extrapolated 
critical times $T_c(\nu)$ defined by
\EQL{eq:TcDt} T_c(\nu)=
\bigl(t_x-t_\Gamma B_x/B_\nu(t_\Gamma)\bigr)/\left(1-B_x/B_\nu(t_\Gamma)\right) \EN
where $B_x=B_\nu(t_x)=\bigl(\sqrt{\nu}Z(t_x)\bigr)^{-1/2}$

Physically, for the Q-trefoils $t_\Gamma$ is roughly when negative $h<0$ helicity density
first appears
and for the anti-parallel cases in section \ref{sec:antip}, $t_\Gamma$ is when an exchange of
the circulation $\Gamma$ between the vortices begins and is why this early time is designated as
$t_\Gamma$. 

Using these $T_c(\nu)$, $t_x$ and $B_x$, figure \ref{fig:QHdRisnuZtime} plots the
following rescaled enstrophy 
\EQL{eq:isnuZBxtime} (T_c(\nu)-t_x)\bigl(B_\nu(t)-B_x\bigr)=
(T_c(\nu)-t_x)\left(\frac{1}{(\sqrt{\nu}Z(t))^{1/2}}-B_x \right)\,. \EN
For the trefoils, the collapse to this form develops at early times compared to $t_x$,
even though the rescaling was chosen empirically and does not obey the time scaling 
implied by the Leray analysis, as discussed in section \ref{sec:Leray}. 
A side benefit of the collapse to early times given by \eqref{eq:isnuZBxtime} is that this 
justifies making a connection between the very small viscosity, early time analysis of section 
\ref{sec:regularity} to the crossing of the $\sqrt{\nu}Z(t)$ at $t=t_x=40$ in figure \ref{fig:QsnuZ}a.
\DEL{It is empirical because it is not completely clear when the \eqref{eq:isnuZBxtime} scaling 
begins, which influences the determination of the $T_c(\nu)$ using \eqref{eq:TcDt}.}

\section{Anti-parallel reconnection: Circulation exchange and $\sqrt{\nu}Z$ collapse 
\label{sec:antip}}

\begin{figure}
\bminic{0.45}\hspace{-0mm}
\includegraphics[scale=.18]{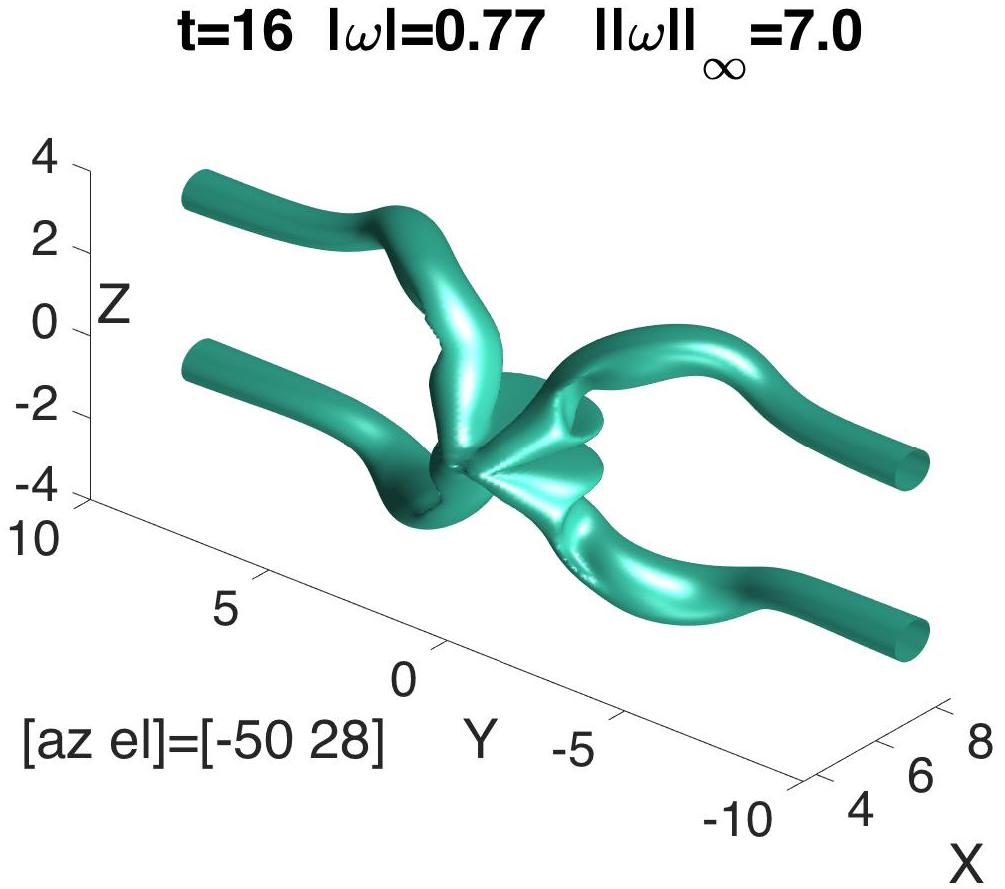} ~~~~~
\emini~~~\bminic{0.45}
\includegraphics[scale=.18]{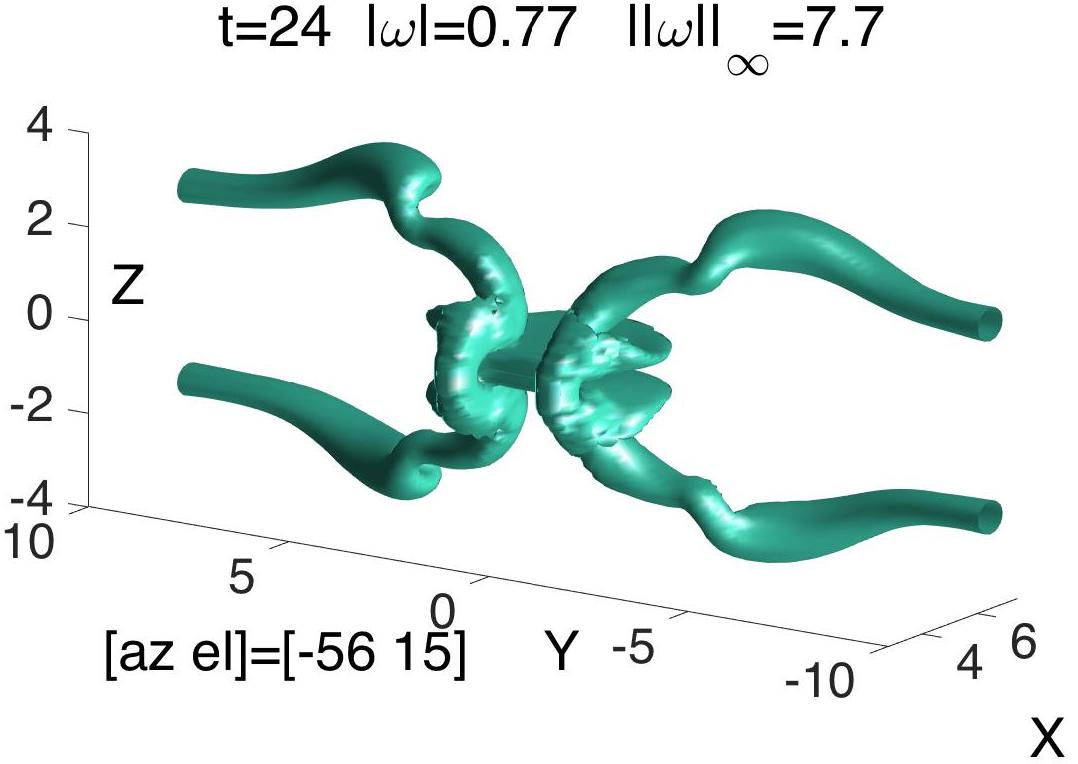} 
\emini
\caption{\label{fig:antiPt16-24} Anti-parallel vortex isosurfaces at $t=16\approx t_\Gamma=16.5$ 
and $t=24\approx t_x$.  Reconnection is just beginning at $t=16$ and the structure 
is nearly identical to the final stage in anti-parallel Euler calculations \citep{Kerr2013b}. 
The period $t_\Gamma=16.5\leq t\leq t_x\approx25.5$ covers the first reconnection, 
with the $t=24$ frame showing how the heads reconnect across the $z=0$ plane to form 
vertical $\omega_z$ bridges with gaps in the original horizontal
$\omega_y$, whose remnants on the $y=0$ symmetry plane are concentrated in two
vortex sheets.}
\end{figure}

The second set of calculations showing $(\sqrt{\nu}Z(t))^{-1/2}$ scaling are new anti-parallel 
calculations using the same initial state as in \cite{Kerr2013a}, 
but higher Reynolds numbers, shorter $y$-domains 
and larger $z$-domains.  As then, the circulation of the initial $y$-vorticity is 
$\Gamma_y\approx5$ and the separation of the vortices is $\delta_a=2a=4$ with $t_a=3.2$.
Figure \ref{fig:antiPt16-24} shows the full periodic vortex structures as the first reconnection 
begins at $t=16$ and ends at $24$.

Two advantages of the anti-parallel configuration for studying the beginning of reconnection 
are that more resolution can be applied to the reconnection zone and how the components of 
vorticity are attached to one another can be easily identified.
The disadvantages are that due to symmetries the global helicity is identically zero and 
the initial integral norms increase as the domain is increased, making comparisons to 
the mathematics problematic.

\begin{figure} \bminil{0.72}
\includegraphics[scale=0.25]{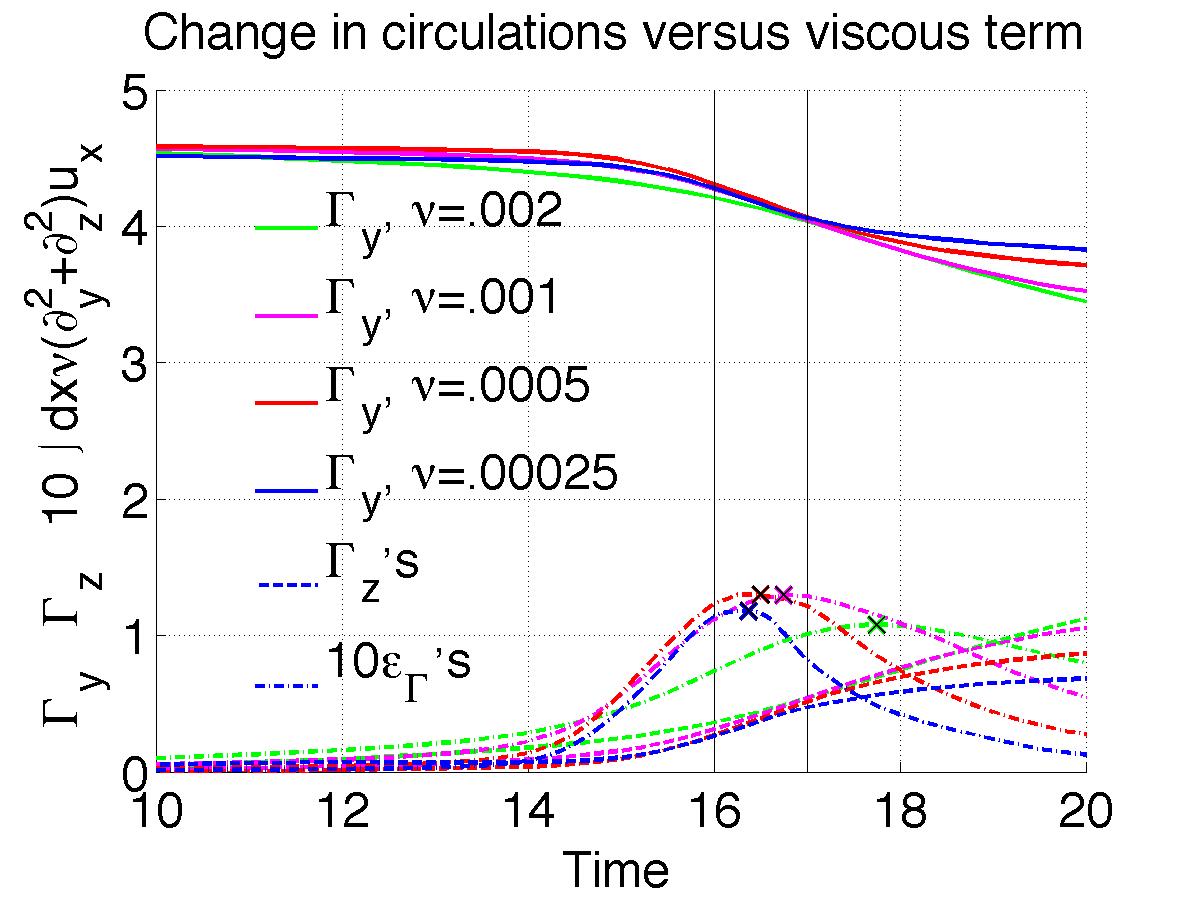}
\emini~~~~ \bminir{0.25}\vspace{-5mm}
\caption{\label{fig:GammasepsG} Anti-parallel circulation exchange.
Between $t=14$ and 18 there is a small, but finite, viscous exchange of circulation between the 
two symmetry planes. The rate of change $\epsilon_\Gamma$ can be predicted exactly by 
a line integral \eqref{eq:dGammadt} where the 
$y=0$ and $z=0$ symmetry planes meet and is plotted for each case. 
The time of the maximum of this viscous exchange retreats back towards $t=16$.
}\emini \\
\bminil{0.72} 
\includegraphics[scale=0.25]{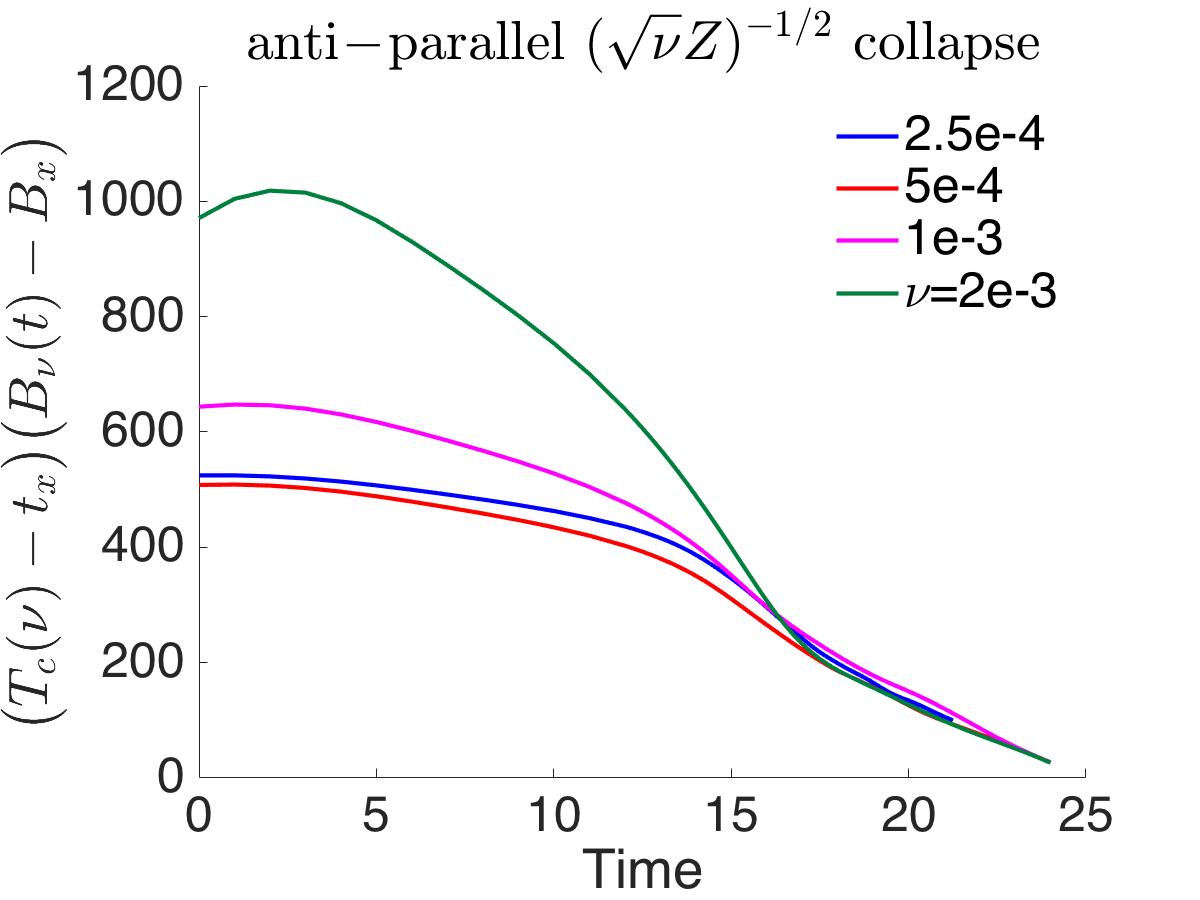}
\emini~~~~ \bminir{0.25}\vspace{-5mm}
\caption{\label{fig:dNSsnuZ} Evolution of the inverse, scaled enstrophy for the new
anti-parallel calculations using \eqref{eq:isnuZBxtime}
with $B_\nu(t)=(\sqrt{\nu}Z(t))^{-1/2}$ \eqref{eq:TcDt} and $B_x=B_\nu(t_x)$.
This scaling regime begins with the brief self-similar exchange of circulation in 
figure \ref{fig:GammasepsG} at $t_\Gamma\lesssim16.5$. 
The three-dimensional structure of the reconnecting vortices at later
times is given in \cite{Kerr2013a}.
}\emini 
\end{figure}

Between the beginning and end of reconnection in figure \ref{fig:antiPt16-24}, 
figure \ref{fig:dNSsnuZ} shows collapse using \eqref{eq:isnuZBxtime} of $B_\nu(t)$ \eqref{eq:Bnu} 
similar to that shown for the trefoil in figure \ref{fig:QHdRisnuZtime}.  
The anti-parallel domain size, in these cases just in $z$, must also be
increased for the smaller viscosities to see the $B_\nu(t)$ collapse.  What the anti-parallel 
cases can do in figure \ref{fig:GammasepsG}, that the trefoil cases have not yet done,
is show how the $\sqrt{\nu}Z(t)$ scaling begins. This is with a spurt of 
$\nu$-independent circulation exchange between the reconnecting vortices. 
The properties plotted are the $y=0$ and $z=0$ symmetry-plane circulations, 
$\Gamma_y(t)$ and $\Gamma_z(t)$, defined as:
\EQL{eq:Gammayz}\begin{array}{rl}
\medskip\Gamma_y(t)=\int_{y=0}\omega_y dx\,dz,~ & \mbox{the integral of}~
\omega_y~\mbox{on the $x\!-\!z$ perturbation plane.} \\
\Gamma_z(t)=\int_{z=0} \omega_z dx\,dy,~ &\mbox{the integral of}~
\omega_z~\mbox{on a $x-y$ dividing, symmetry plane,} 
\end{array} \EN
and $\epsilon_\Gamma(t)$, the viscous exchange of circulation between the symmetry plane.
This is defined as this integral along the $y=z=0$ $x$-line where the two symmetry planes meet:
\EQL{eq:dGammadt} \epsilon_\Gamma(t)=
\nu\int dx \left(\frac{\p^2}{\p_y^2}+\frac{\p^2}{\p_z^2}\right)u_x(t)
=\ddto{\Gamma_y}(t)=-\ddto{\Gamma_z}(t) \,. \EN
The line integrals $x\in{\cal C}_j$, $j=y$ or $z$, \eqref{eq:Gamma} for $\Gamma_y(t)$ 
and $\Gamma_z(t)$ follow the perimeters of the two symmetry planes.

\ITM\item Note that because $d\Gamma_y/dt=-d\Gamma_z/dt=\epsilon_\Gamma$ exactly,
the sum $\Gamma_y(t)+\Gamma_z(t)=\Gamma_y(t=0)$ is constant during this process.
\ITN

What figure \ref{fig:GammasepsG} shows is that at a $\nu$-independent time of 
$t_\Gamma\approx16.5$, the depletion of the original circulations and the generation of 
reconnected circulations all cross as the exchange rates $\epsilon_\Gamma$
collapse onto a common curve for the three smallest viscosities.

Two immediate effects of this finite circulation exchange are shown by 
the $t=16$ vorticity isosurfaces of figure \ref{fig:antiPt16-24}. 
The formation of significant isosurface of nearly 
vertical reconnected $\omega_z$ and the collapse of the remaining $\omega_y$ from 
the initial state into a horizontal vortex sheet. 
The $t=24$ frame shows how this first reconnection ends, with
almost all of circulation transferred into the new vortices and the formation of
twist and helicity along the vortices. 

How are the structures at these two times related to the times covered by the linearly
decreasing self-similar collapse \eqref{eq:isnuZBxtime} in figure \ref{fig:dNSsnuZ}? 
The first time, designated $t_\Gamma=16$ 
is when both the exchange of circulation $\Gamma$ during reconnection and the self-similar collapse 
begin, and the second time, $t=24\approx t_x=25$ is when both the reconnection and self-similar
collapse end.

The evidence is that by using $t_\Gamma=15$ and $t_x=40$ for the Q-trefoil that
 $t_\Gamma\leq t\leq t_x$ is also the period for self-similar $B_\nu(t)$ collapse, as discussed in 
section \ref{sec:SS}, and for the period of first reconnection, to be discussed in 
section \ref{sec:reconnection}. 

What are $t_\Gamma$ and $t_x$ for the Q-trefoil and anti-parallel in terms of their respective $t_a$, 
the nonlinear timescales  \eqref{eq:ta} based upon the separations of their reconnecting strands? 
For the anti-parallel cases, one can choose $t_\Gamma=16.5\approx5t_a$ to represent the maximum 
of the circulation exchange due to reconnection.  For the Q-trefoils, $t_\Gamma=15=7.5t_a$, 
based upon when linearly decreasing $B_\nu(t)$ \eqref{eq:Bnu} appears in figure 
\ref{fig:QHdRisnuZtime} and the first appearance of negative helicity, 
as discussed in section \ref{sec:Hnegative}. In terms of $t_a$, 
the trefoil's helicity might be delaying the start of reconnection.

For $t_x$ determined by either when their respective $B_\nu(t)$ \eqref{eq:Bnu} cross or when gaps 
appear in their respective vorticity isosurfaces, for the anti-parallel cases $t_x\approx 25=7.8 t_a$
and for the Q-trefoil, $t_x=40=20t_a$
This shows that the helicity slows the trefoil reconnection process considerably once it has 
started and it was the long duration of this phase that first indicated the possibility of the 
new scaling about $t=t_x$ in a way that the anti-parallel calculations did not.

\subsection{Leray similarity and extensions. \label{sec:Leray}}

For the line integral that determines $\epsilon_\Gamma(t)$ \eqref{eq:dGammadt} 
to remain finite as the viscosity decreases,
the second derivative of the velocity terms $(\p_y^2+\p_z^2)u_x$ must grow in a singular manner.  
How it grows can be understood in terms of the \cite{Necasetal96} extension of the original
similarity proposal of \cite{Leray34}.

Given the Navier-Stokes equation \eqref{eq:NS}, \cite{Leray34} proposed the following 
scaling for solutions of the Navier-Stokes equation:
\EQL{eq:Lerayform} \bu(\bx,t)=\frac{\Gamma_L}{\delta_\nu(t)}\bff(\by)
=\frac{1}{\delta_\nu(t)}\bGamma(\by)\quad{\rm or}\quad \bu(\bx,t)=U_\nu(t)\bff(\by) \EN
where $\delta_\nu(t)\sim\sqrt{2a_L(T-t)}$ is a collapsing length scale,
$\by=\bx/\delta_\nu(t)$, $U_\nu(t)=\Gamma_L/\delta_\nu(t)\sim (\Gamma_L/\sqrt{2a_L})(T-t)^{-1/2}$
and $\bGamma(t)=\Gamma_L\bff(\by)$ sets the spatial structure.  
The scalars $a_L$ and $\Gamma_L$ have the same units as the viscosity and circulation. 
Inserting \eqref{eq:Lerayform} in \eqref{eq:NS} one gets 
\EQL{eq:NSLeray}
\begin{array}{r}-\nu\Delta\bGamma+a_L\bGamma+a_L(\by\nabla\bGamma)
+(\bGamma\bdot\nabla)\bGamma +\nabla P=0 \\
\nabla\bdot\bGamma=0 \end{array}\Bigr\}\quad  {\rm in}~ \bbR^3 \EN
The similarity predictions of some aspects the new $\sqrt{\nu}Z(t)$ regime
and whether these are observed, even if only briefly, are now
considered.  The predictions are obtained by multiplying 
$U_\nu(t)=\Gamma_L/\delta_\nu(t)$ and $\delta_\nu(t)$ in the appropriate dimensional 
combinations. To begin, the similarity Leray prediction for $\sbl{\omega}{\infty}{}$ is 
\EQL{eq:ominftyLeray} 
\sbl{\omega}{\infty}{}\sim U_\nu(t)\nabla\cdot(\bx/\delta_\nu)f’
\sim U_\nu(t)/\delta_\nu(t)=\frac{\Gamma_L}{\nu(T_c-t)}\,. \EN
This is not observed except for $t<20$ when the calculations are nearly Euler.  That is,
Navier-Stokes $1/\sbl{\omega}{\infty}{}$ is briefly linear in time, but 
figure \ref{fig:EulerNSomaxtsmall} shows that the
Navier-Stokes $\sbl{\omega}{\infty}{}$ are all bounded by the non-singular growth of the 
Euler $\sbl{\omega}{\infty}{}$ for $t>20$. This is discussed further in section 
\ref{sec:regularity}. 

What is the Leray similarity prediction for the cubic velocity norm $L^{(3)}_\ell$
\eqref{eq:LpHsT3R3}? Using \eqref{eq:Lerayform} one gets 
\EQL{eq:L3} L^{(3)}=L^{(3)}_{\ell=\infty}=\sbl{u}{L^3(\bbR^3)}{}\sim 
\left(U_\nu^3(t)\delta_\nu^3(t)\right)^{1/3}\sim{\cal O}(1)\,. \EN 
This prediction of constant $L^{(3)}$ 
is consistent with figure \ref{fig:HL3H12}c which shows that for two of the trefoil calculations, 
$L^{(3)}_\ell$ is remarkably independent of both time and $\nu$ and 
decreases less rapidly than the kinetic energy.  However, \cite{EscauSS03} has shown
that bounded $L^{(3)}=\sbl{u}{L^3}{}$ is the most refined restriction against singularities of the
Navier-Stokes equation in a proof by contradiction that uses the scaling in \eqref{eq:L3},
despite the singular assumption in \eqref{eq:Lerayform}.

One way to resolve that apparent contradiction is if one can assume that there are very large 
effective singular times $T$ such that transient, singular growth could be allowed for some norms,
so long as the transients end long before $t=T$ is ever reached.  
This would be consistent with the very brief Leray growth 
of $\epsilon_\Gamma$ in figure \ref{fig:GammasepsG}.  For times after that
spurt in circulation exchange, it seems plausible that Leray similarity might set the scale
for those norms, but their subsequent time dependence would not obey the Leray similarity 
prediction. The use of the linearly extrapolated singular
times in the self-similar collapse \eqref{eq:isnuZBxtime} of $B_\nu(t)$ 
in figures \ref{fig:QHdRisnuZtime} and \ref{fig:dNSsnuZ} reflects that point of view.
\ITM\item For those reasons, only the viscosity dependencies will be given
for the remaining properties with elements of Leray scaling. \ITN

For the circulation exchange in \eqref{eq:dGammadt}, \cite{Necasetal96}
provides an estimate the second velocity derivatives in \eqref{eq:dGammadt} where
\EQL{eq:ppu} \left(\p_x^2+\p_y^2\right)u \sim U_\nu/\delta_\nu^2 \sim \nu^{-3/2} \,.\EN
Then the Leray scaling for line integral over $\delta_\nu(t)$ would be 
$\epsilon_\Gamma\sim (\nu \delta_\nu)(\nu^{-3/2})\sim {\cal O}(1)$, as observed 
in figure \ref{fig:GammasepsG}.

Finally, let us consider $\sqrt{\nu} Z$. The $\nu$ dependence of its Leray scaling
would be 
\EQL{eq:snuZLeray} \sqrt{\nu} Z\sim \nu^{1/2} \delta_\nu^3 (U_\nu/\delta_\nu)^2\sim {\cal O}(1)
\quad{\rm or}\quad Z\sim \nu^{-1/2} \,,\EN
as observed at $t=t_x$ in figures \ref{fig:QHdRisnuZtime} and \ref{fig:dNSsnuZ}.

\section{Evolution the trefoil's topology during reconnection \label{sec:reconnection}}
\begin{figure} 
\includegraphics[scale=0.36]{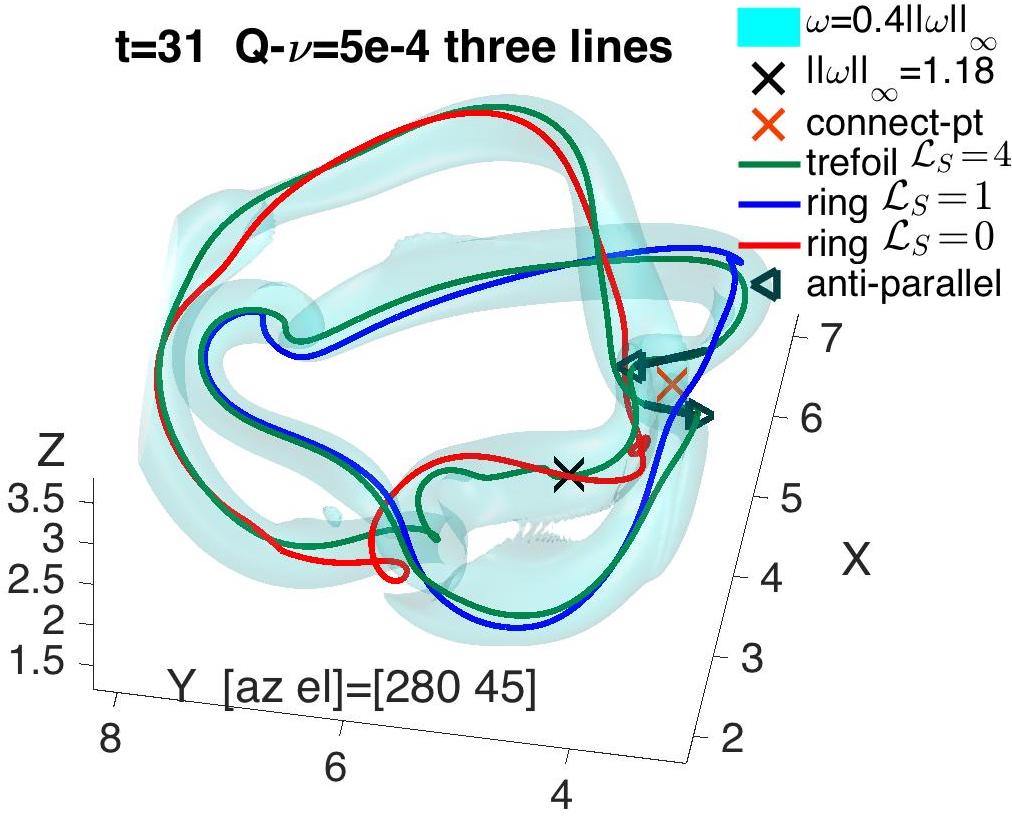}
\caption{\label{fig:T31} A single vorticity isosurface plus three closed vortex lines 
at $t=31$, the first time that visible reconnection is observed. 
The green trajectory follows a trefoil trajectory seeded near $\sbl{\omega}{\infty}{}$, 
indicated by {\bf X}.  The green trajectory's self-linking is ${\cal L}_S=4$, which can be 
split into $W\!r+T\!w=2.85+1.15=4$ At the closest approach of the trefoil's two loops, 
due to an extra twist, the loops are anti-parallel, as indicated by two arrows.
Between them is the {\it reconnection zone} whose mid-point is shown by the orange cross.  
The red ${\cal L}_{Sr}=0$ and blue ${\cal L}_{Sb}=1$ linked trajectories were seeded 
on either side of this point, in the direction perpendicular to the loops' separation line.
Their total linking is ${\cal L}_t=2{\cal L}_{rb}+{\cal L}_{Sr}+{\cal L}_{Sb}=2+0+1=3$, 
the linking of the original trefoil.}
\end{figure}

\begin{figure}
\includegraphics[scale=0.36]{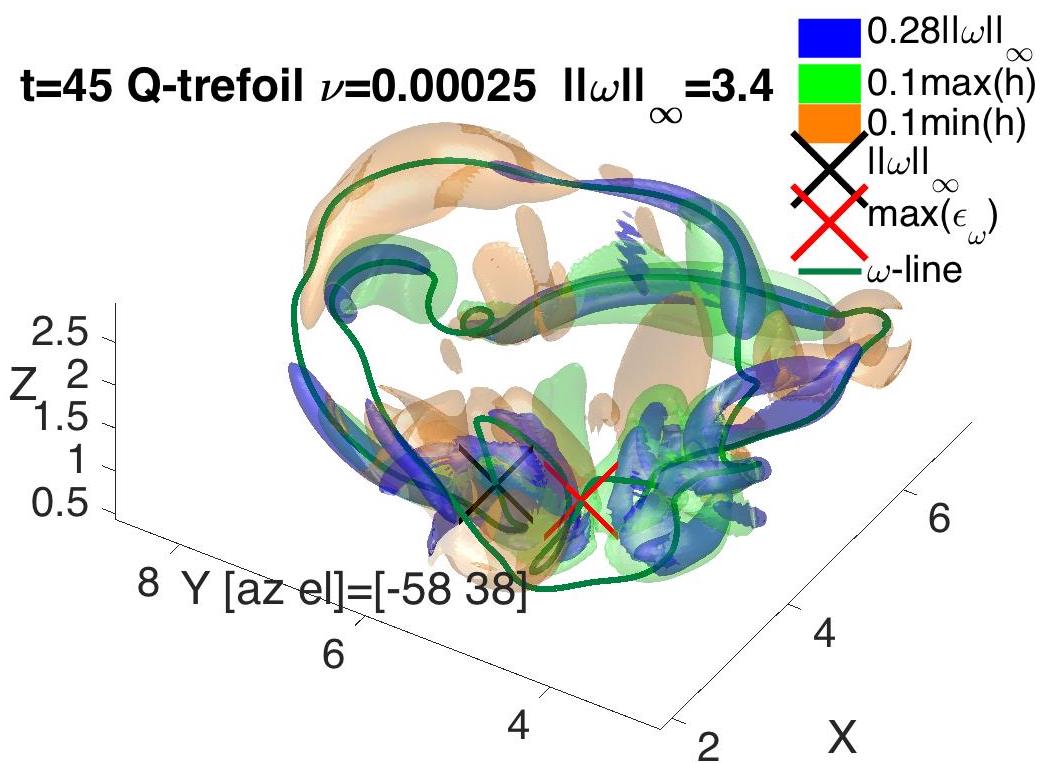}
\caption{\label{fig:T45} Isosurfaces and one vortex line at $t=45$ just after the first
reconnection has ended.  Vorticity isosurfaces are in blue and 
the helicity isosurfaces are at $0.1\max(h)$ in green and
$0.1\min(h)$ in yellow, where $\max(h)=0.62$ and $\min(h)=-0.26$. 
A gap without strong vorticity, but twisted and bent vortices to either side, 
now covers the {\it reconnection zone} to the right of red {\bf X}.
Nonetheless, except in that zone, the vortex line seeded at the point of maximum vorticity 
at {\bf X} still has the flavour of the
original trefoil as it circumnavigates the centre twice and passes through regions 
with large vorticity and large helicity of both signs.  Green positive helicity overlying
twisted blue vorticity dominates to the right of the gap.  The region between the 
black {\bf X} and red {\bf X}, to the left of where reconnection began at $t=31$, 
is now dominated by the negative helicity discussed in section \ref{sec:Hnegative} along with
the additional regions of large negative helicity outside the {\it reconnection zone}.
} 
\end{figure}

The purpose of this section is to outline the temporal changes to the three-dimensional 
structure from the beginning of observable reconnection until just after reconnection ends
using two times. 
The position of $\sbl{\omega}{\infty}{}$ is the black {\bf X} and the {\it reconnection zone}
is indicated by a orange/red {\bf X}.  The definition of the {\it reconnection zone} changes with 
time, but the relative locations of these points with respect to the overall, slowly rotating trefoil 
structure does not change from where they sat at $t=6$ in figure \ref{fig:T6}.
Figure \ref{fig:T31} at $t=31$ was chosen to represent the beginning of reconnection because it 
was the first time that linked rings, rings whose trajectories originated within the trefoil's 
{\it reconnection zone}, could be identified.
Figure \ref{fig:T45} at $t=45$ was chosen to represent the phase just after reconnection 
has ended because this is the first time that the reconnection has created a clear gap in the
the blue vorticity isosurface. 

Both figures show one mid-level vorticity isosurface and one green trefoil vortex trajectory with
additional diagnostics in each figure highlighting those features that are particularly important 
for that phase of the evolution. 
The trefoil trajectories originate at the {\bf X}, or passes this point, and both
circumnavigate the central $z$ axis twice before closing almost exactly upon themselves. 
The perspectives are tilted so that the overall trefoil structure can be seen with
the zone between the red/orange and black {\bf X}'s in the foreview, a zone
that slowly rotates from right to left along with the entire trefoil. 


How reconnection begins is shown in figure \ref{fig:T31} at $t=31$ using 
two additional linked single vortex loops in red and blue that 
were seeded on opposite sides of the orange {\bf X}, which is the mid-point 
between the closest approach of the green trefoil's two loops. Due to an acquired twist in the 
green trefoil, these segments of the loops are anti-parallel about the orange {\bf X}, 
as indicated by the arrows, and define the {\it reconnection zone} because this is where
anti-parallel, helicity preserving reconnection begins, as discussed further in section 
\ref{sec:Hnegative}.  Due to this twist, by
using \eqref{eq:Gauss}, the total self-linking number \eqref{eq:selflink} of the green trajectory 
is ${\cal L}_{Sg}=4$.  Since the 
red and blue loops are linked and the blue loop has twist+writhe whose self-linking 
is ${\cal L}_{Sb}=1$, the total linking number of the red and blue loops is 
${\cal L}_t={\cal L}_{rb}+{\cal L}_{br}+{\cal L}_{Sb}=3$ using \eqref{eq:Hlink} with $\Gamma=1$,
equal to the total linking number of the original trefoil. This demonstrates why,
if helicity is simply ${\cal H}=\Gamma^2{\cal L}$ \eqref{eq:Hlink}, reconnection by itself 
need not result in a change in the total helicity \citep{Laingetal2015}. 
However, for the continuum Navier-Stokes equations, it is not that simple, as shown by
the self-linking of the green trefoil with ${\cal L}_{Sg}=4\neq3$. 

When does the reconnection finish?  Figure \ref{fig:T45} at $t=45$ tells us that 
this is before $t=45$ based upon the {\it reconnection zone} gap that has formed 
in the blue vorticity isosurfaces. This is consistent with the reconnection ending 
at the $t=t_x\approx40$ convergence of $\sqrt{\nu}Z(t)$ in figure \ref{fig:QsnuZ}.
This gap is to the right of the position of $\max(\epsilon_\omega)$ at the red {\bf X} and
unlike in figure \ref{fig:T31} at $t=31$, the green trefoil trajectory in figure \ref{fig:T45} 
at $t=45$ goes around this gap, not through it. In addition, the sign and colour of the 
helicity density $h$ isosurfaces change across the gap. 
To the right of the gap, positive green helicity surrounds and covers
the twisted blue vorticity isosurfaces 
and to the left of the gap, near the {\bf X} with 
$\sbl{\omega}{\infty}{}$, negative yellow/orange helicity dominates.

By all of these measures, it is the formation of this gap at $t\gtrsim40=t_x$
that marks the end of the first
reconnection and is the most relevant timescale for comparing with the experiments 
in the next section. Furthermore, the new twists around this gap mark
the beginning of a new phase of even stronger enstrophy growth that 
leads to the development of finite dissipation $\epsilon=\nu Z$ at $t_\epsilon\approx 2t_x$ in
figure \ref{fig:QsnuZ}b. This persists over the finite time interval that follows $t_\epsilon$
and would be consistent with the formation of a {\it dissipation anomaly}, 
that is a finite viscosity-independent loss of the total energy in a finite time.  

\section{Helicity \label{sec:helicity}}

\begin{figure} 
\includegraphics[scale=0.32]{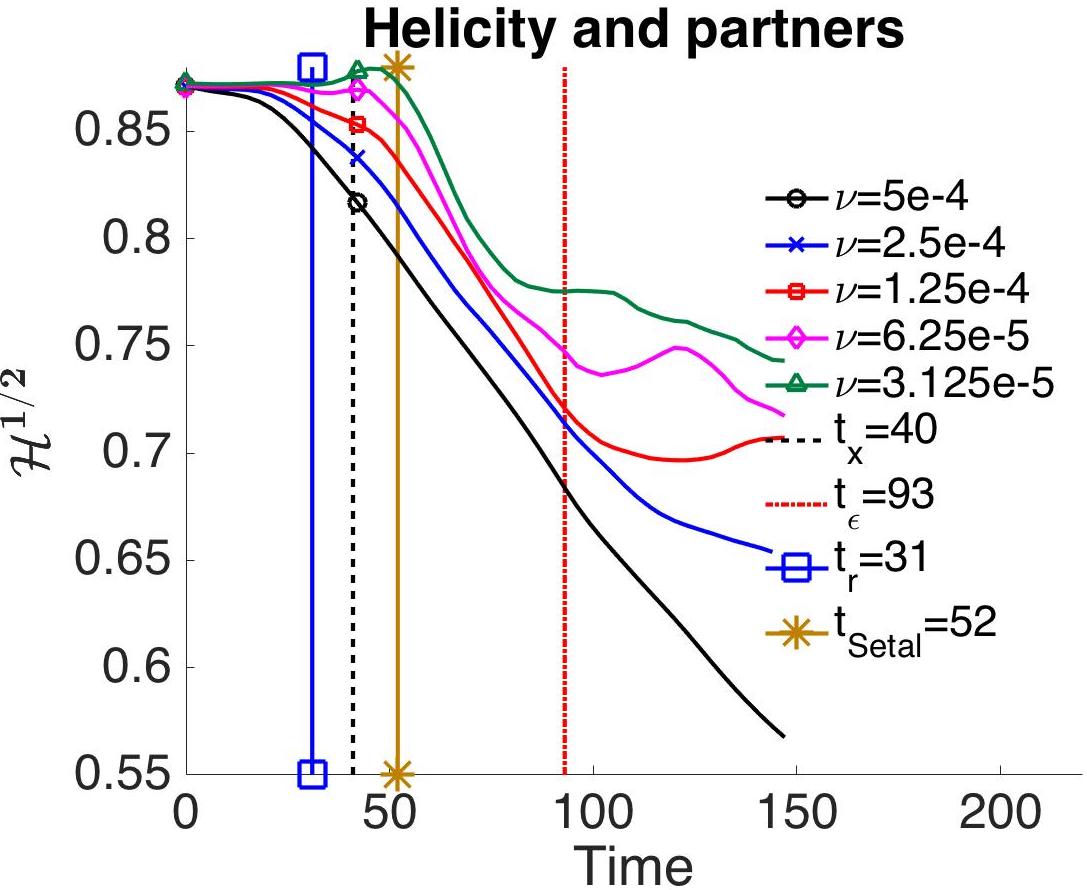}
\begin{picture}(0,0) \put(70,250){{\LARGE\bf (a)}}\end{picture} \\
\includegraphics[scale=0.32]{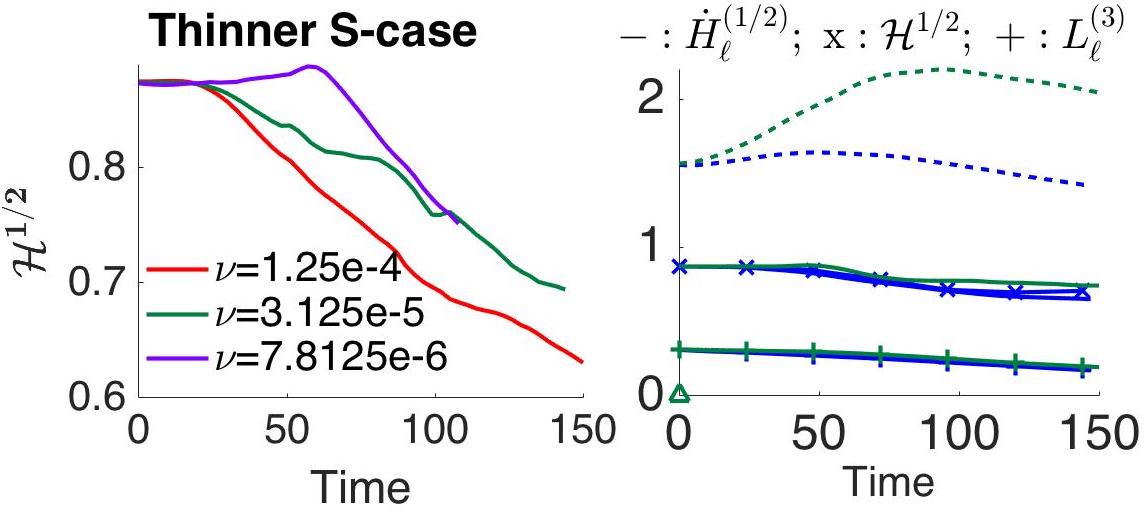}
\begin{picture}(0,0)\put(-235,100){{\LARGE\bf (b)}} 
\put(-80,80){{\LARGE\bf (c)}}\end{picture} 
\caption{\label{fig:HL3H12} {\bf a:} Time evolution of the helicity
${\cal H}^{1/2}$ and its partners. In the main frame are 5 viscosities for the Q-trefoils: 
$\nu=0.0005$ to $\nu=0.00003125$.  Also given are important times during the evolution of the 
trefoil.  The blue-$\square$ line is $t=31$, the time when the first signs of reconnection are 
visible as in figure \ref{fig:T31}. The dashed line is $t_x=40$, the time when 
all the $\sqrt{\nu}Z$ meet and has been designated the reconnection time.  
The brown-$\star$ line is $t_{Setal}=52$, roughly the equivalent time to 
when the \cite{ScheeleretalIrvine2014a} experiment probably ends. The red-dot-dash line is the 
time $t_\epsilon=93$ when the dissipations $\epsilon=\nu Z$ in figure \ref{fig:QsnuZ} 
reach a common value.  
{\bf b:} ${\cal H}^{1/2}$ for S-trefoils with thinner core radius $r_e$ 
at three viscosities, demonstrating that once the viscosity is small enough that
after a long period of helicity preservation, the time when helicity decay begins 
$t\approx1.25 t_x\approx50$ is independent of the difference in their core thicknesses.
{\bf c:} Comparison of global helicity ${\cal H}$ and two regularity diagnostics.
From top to bottom, $\dot{H}^{(1/2)}_\ell =\sbl{u}{\dot{H}^{1/2}(\bbT_\ell^3)}{}$ 
\eqref{eq:LpHsT3R3b} global helicity ${\cal H}^{1/2}$ \eqref{eq:helicity}
and $L^{(3)}_\ell=\sbl{u}{L^3(\bbT_\ell^3)}{}$ \eqref{eq:LpHsT3R3} 
for the $\nu=2.5$e-4 and $\nu=3.125$e-5 Q-trefoils. 
All three diagnostics have been normalised to have the units of circulation and
only $\dot{H}^{(1/2)}_\ell$ shows a dependence upon the viscosity. }
\end{figure}
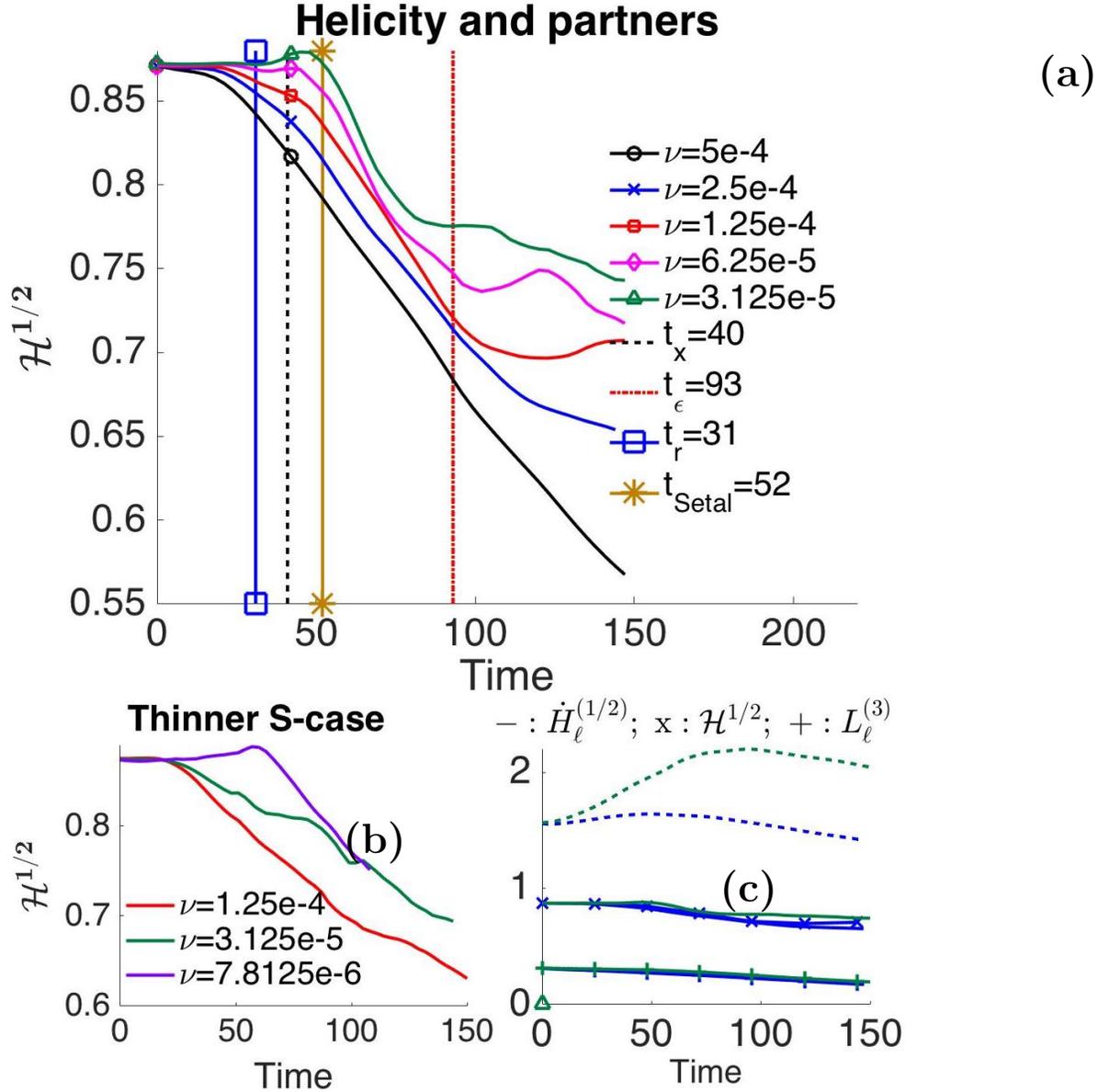 

While the focus of this paper has been on comparisons of vorticity norms 
to mathematical bounds, the inspiration for this study came from 
the unexpected experimental claim that the centre-line helicity of the vortex knots in
\cite{ScheeleretalIrvine2014a} was preserved through reconnection events and for 
one trefoil case was preserved until the experiment ended.  

These experimental trefoil vortices and linked vortex rings were generated and observed by 
yanking 3D-printed knots, shaped into hydrofoil ribbons, out of a water tank, 
as described by \cite{KlecknerIrvine2013}.  
The ribbons were covered with hydrogen bubbles that were shed as 
the low pressure vortex cores were generated, providing a means for observing those 
cores.  The hydrogen bubble filaments that form within these cores were then used to define 
the centre-line helicity diagnostics of \cite{ScheeleretalIrvine2014a}. 
Qualitative evidence for when the topology changes
was provided by consecutive time-frames when the bubbles on these filaments disperse, then reform.

Many questions have been raised about their unconventional diagnostics, so a set of 
simulations that could either confirm or explain their results would help.  
To make those comparisons, common definitions of the nonlinear 
and reconnection timescales, $t_f$ \eqref{eq:ta}
and $t_x$ are needed, where figure \ref{fig:QsnuZ} defines $t_x$ for the simulations.
Ideally, the definitions of both $t_f$ and $t_x$ should be related to
some aspect of the trajectories of the observed vortex lines because this 
is the only diagnostic provided by the experiments. What properties might
the experimental trajectories and the vortex lines of the simulations share? 
Due to the intricate and under-resolved structures that form during reconnection in
both the experiments and simulations, looking for common properties 
at the reconnection time has been difficult.

\subsection{Experimental reconnection and helicity depletion timescales \label{sec:Xtimes}}

There are two routes for estimating the experimental timescales.
\ITM\item[1.)] An estimate for the nonlinear timescale $t_f$ \eqref{eq:ta} can be 
made using the estimated circulations $\Gamma$ of the shed vortices 
and the radii of the knots $r_f$ \eqref{eq:trefoil}.  
While the radii can easily be determined for both the experiments and these simulations, 
the circulations $\Gamma$ given for the experiments were not measured, but 
estimated based upon the flat plate estimate for how flow over the hydrofoil generates
$\Gamma$. This approximation neglects the camber of the trefoil 
ribbon. Therefore, only the order of magnitudes of the scaled times $t/t_f$ between 
the simulations and the experiments can be compared. 
\item[2.)] A visual timescale for $t_x$ can be determined from the experimental reconnection 
figures by first identifying the first clear, visible gap in the trefoil structure. Since
this is after the reconnection ends, a better estimate of the reconnection times $t_x$ for
the experiments is the first frame before the clear gap with evidence that 
the topology is changing. That is consecutive frames when the bubble lines disperse, then reform.
\ITN

For these simulations, the clear gap in figure \ref{fig:T45} at $t=45$ is consistent
with the reconnection being after the inferred end of reconnection at $t_x\approx40$.
For the largest trefoil from \cite{KlecknerIrvine2013}, with
a camber correction applied to the flat-plate method for estimating the circulation
$\Gamma$, the two methods for estimating $t_x$ give
consistent values, with a reconnection time of $t_x\approx350$ms. 
Comparisons between Q-trefoil figures at $t=36$, 40 and 45 and the $t=300$, 350 and 400 figures
from \cite{KlecknerIrvine2013} show the same stages in their development and are the topic of
another paper \citep{KerrFDR17}.

Detailed comparisons with the timescales and movie from \cite{ScheeleretalIrvine2014b} 
will require detailed graphics using the thinner S-trefoils, for which $t_x\approx40$
and whose ${\cal H}(t)$ for several $\nu$ are given in figure \ref{fig:HL3H12}b.
For now, the visual evidence indicates that its reconnection gap is 
in the $t=658$ms frame, and looking earlier in the movie, the reconnection would then
be at $t_x\approx638$ms. This is the basis for the $t_{\rm Setal}=52$
line in figure \ref{fig:HL3H12} indicating when that experiment ends. 


The conclusion is that even if figure \ref{fig:HL3H12} does not show helicity preservation 
for all time, it does show that the true helicity ${\cal H}$ \eqref{eq:helicity} 
can be preserved through, and a bit after, the end of first reconnection at $t=t_x\approx40$.  
That is, helicity does not decay at all until $t\approx1.25t_x$ and does
not decay significantly until $t\gtrsim 1.5t_x$, showing consistency with the preservation of the
experimental centre-line helicity.  It also supports the conclusion
that these simulations are representing a physical flow, irrespective of 
whether any of the $\nu\rightarrow0$ limits being proposed become accepted.

The similarities in the ${\cal H}$ timescales for the Q and S-trefoils in figure \ref{fig:HL3H12} 
demonstrates that this behaviour is independent of the core thickness $r_e$ once $r_e$ is 
sufficiently small. This is also observed for plots of
$\sqrt{\nu}Z(t)$ and $(T_c(\nu)-t_x)(B_\nu(t)-B(x)$ \eqref{eq:isnuZBxtime} for 
several S-trefoils, including the $\sqrt{\nu}Z(t)$ crossing at $t_x\approx40$.

\subsection{Negative helicity surfaces \label{sec:Hnegative}}

While the experimental bubble filaments might shed light upon the evolution of the global
helicity, they cannot tell us what the sign of the helicity along the filaments is or what 
the helicity is off of the vortex cores. 
The change in the sign, and colour, of the helicity isosurfaces have already been used to 
highlight the gap in the {\it reconnection zone} in figure \ref{fig:T45} at $t=45$ marked by
a red {\bf X}.  What are the relationships between the different signs of the helicity and 
the vorticity isosurfaces outside of the {\it reconnection zone}? In particular,
how do the $h<0$ zones form and what is their role?

The following is a summary of the evolution of negative helicity using figures
\ref{fig:T6}, \ref{fig:T31} and \ref{fig:T45} at $t=6$, 31 and 45. Further details,
with more figures, are in an additional submitted paper \citep{KerrFDR17}. 
To relate to later times, three points are marked along the trefoil centreline vortex in figure 
\ref{fig:T6}. The vorticity maximum $\|\omega\|_\infty$ at the black {\bf X} plus the red and yellow 
{\bf +} points marking the closest approach of the trefoil's loops and where reconnection will begin.

Noticeable $h<0$ forms very early, growing to the left of the red {\bf +} point in figure \ref{fig:T6}
towards the vorticity maximum at the black {\bf X}.  Growth of local negative $h<0$ becomes stronger 
starting at $t_\Gamma=15$ as the linear scaling of $B_\nu(t)=(\sqrt{\nu}Z)^{-1/2}$ begins 
in figure \ref{fig:QHdRisnuZtime}. Up until $t=31$, this zone of $h<0$ continues to grow 
between the now locally anti-parallel loop segments and the position of $\|\omega\|_\infty$.
The knot on the red trajectory just below the orange-{\bf X} reconnection point covers this
zone of $h<0$.  Above the orange {\bf X}, $h>0$ grows as it envelopes the vorticity isosurfaces.
At this stage, the separation of the $h<0$ and $h>0$ zones 
is largely due to the $\omega$-transport term in \eqref{eq:helicity}. 

In the next $t>31$ stage, due to locally anti-parallel reconnection, zones of equal magnitude and 
oppositely signed viscous dissipation form, as predicted by \cite{Laingetal2015}, which
leads to the strong $h<0$ yellow zone to the left of the red {\bf X}, a gap in the vorticity 
isosurfaces in figure \ref{fig:T45} at $t=45$ and a strong green $h>0$ isosurface to the right.
This $h<0$ yellow zone is between, but not on, the sharp bends in the blue vorticity isosurfaces.
The strongest and most distant $h<0$ yellow isosurface 
along the top loop of remaining branch of trefoil vortex and is most likely due to 
the $\omega$-transport term in \eqref{eq:helicity}. 
The $h<0$ zones between this outer loop and the red {\bf X} can all be associated with
velocity advection out of the {\it reconnection zone}.

Note that almost all of the green $h>0$ isosurfaces envelope blue vorticity isosurfaces and
for the zone to the right of the red {\bf X}, there are extra twists in the
blue vorticity isosurfaces within the green $h>0$ isosurface. These twists show how 
smaller vorticity scales, enstrophy and $h>0$ can be generated, without increasing the global helicity
${\cal H}$ due to the generation of $h<0$ elsewhere.




Can this simultaneous growth of positive and negative helicity density $h$, an integral of
the form $\int |h| d\Omega$,  be expressed by
one of the global helicity's two partner norms, $L^{(3)}_\ell$ 
\eqref{eq:LpHsT3R3} and $H^{(1/2)}_\ell$ 
\eqref{eq:LpHsT3R3b}? 

The cubic velocity norm $L^{(3)}_\ell$, as discussed in section \ref{sec:Leray},
is virtually constant and cannot represent the integral growth of $|h|$. However, $H^{(1/2)}_\ell$ 
does increase, although less rapidly than its upper bound of $\sqrt{2EZ}$, and
could represent this growth as $\nu$ decreases as shown in figure \ref{fig:HL3H12}.
Especially when compared with the relatively constant global helicity ${\cal H}^{1/2}$.
This suggests that $H^{(1/2)}_\ell$ might be the best property to investigate
further for what controls (bounds from above) the Navier-Stokes equation, with both
applied analysis and numerically.

\section{Small or zero viscosity at early times. \label{sec:regularity}}

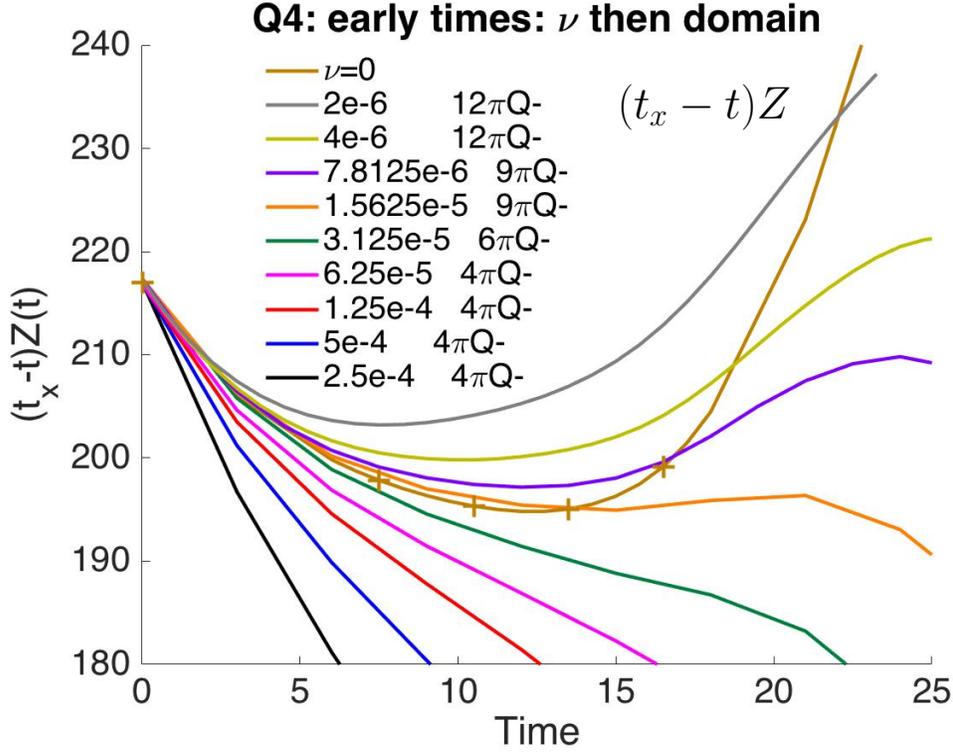
\begin{figure} 
\includegraphics[scale=.32]{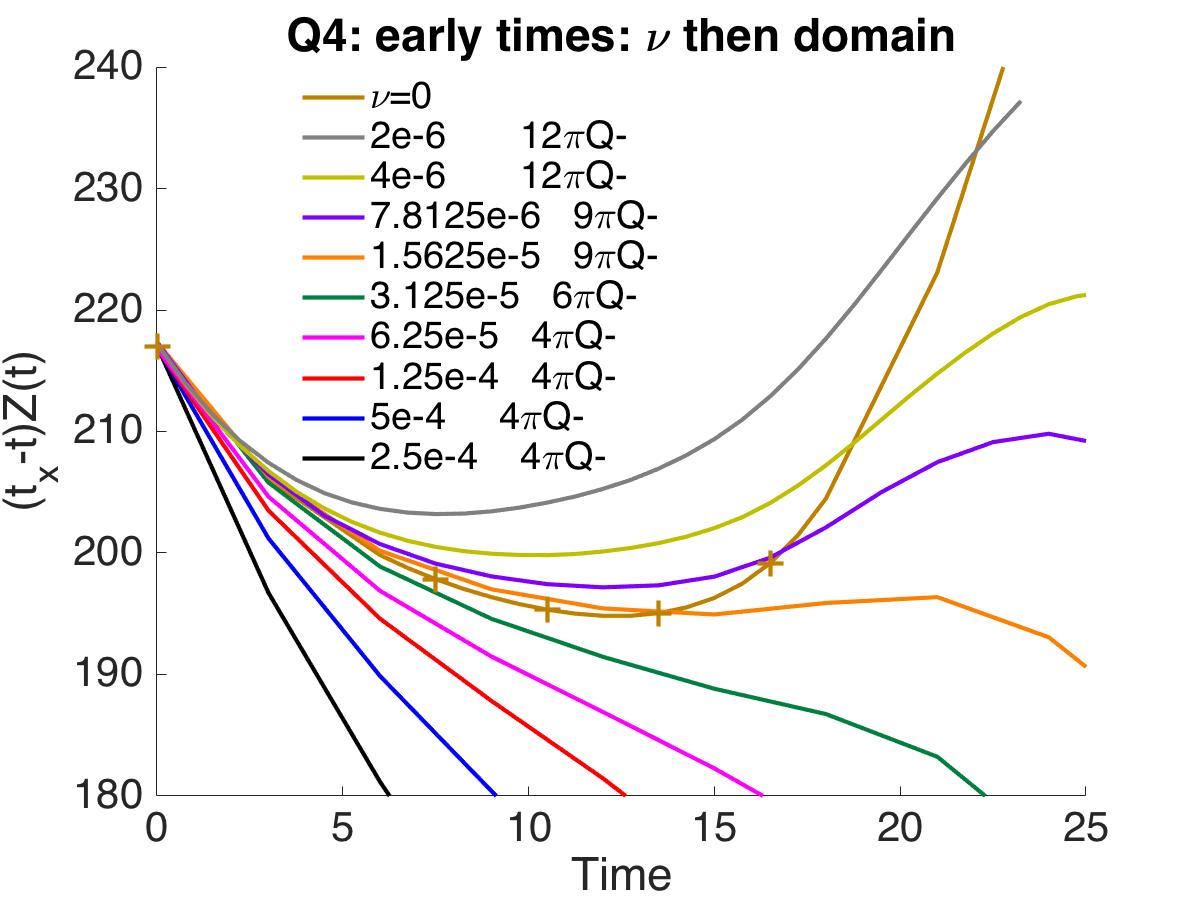}
\begin{picture}(250,0)\put(230,250){\LARGE $(t_x-t)Z$} \end{picture}
\caption{\label{fig:EulerNS-Zttx} 
$(t_x-t)Z$ at early times when the enstrophies of the Euler and very small $\nu$ calculations 
are not affected by under-resolution of $\|\omega\|_\infty$.
The purpose of this figure is to show that as the viscosity continues to decrease,
the Navier-Stokes enstrophies can exceed the Euler enstrophy as late as $t=24$. At later times,
$\sqrt{\nu}Z$ for the smallest viscosity case, $\nu=2$e-6 in a $(12\pi)^3$ domain, joins its continuation in
figure \ref{fig:QsnuZ}.}
\end{figure}

To complement the five $\nu\geq3.125$e-5 cases in figures \ref{fig:QsnuZ} and 
\ref{fig:QHdRisnuZtime}, Euler calculations in several different domains and 
several Navier-Stokes calculations with $\nu\leq1.5625$e-5 have been run. 
While all of the $\nu\geq3.125$e-5 Navier-Stokes calculations are 
resolved in terms of $Z$ through $t=40$, all of the Euler calculations and the additional 
$\nu<3.125$e-5 calculations are under-resolved in terms of $\sbl{\omega}{\infty}{}$ and spectra 
for $t>25$.  Despite this, these additional calculations are reliable at early times and 
therefore can address questions that the $\nu\geq3.125$e-5 calculations cannot.

{\bf Questions} The first goal for the Euler cases was to determine which of these 
calculations would be the most representative of later times run in large domains.  This included 
determining whether $\sbl{\omega}{\infty}{}$ and $Z$ for fine-resolution
calculations would grow in a manner that is consistent with the exponential of exponential 
Euler growth identified in \cite{Kerr2013b}. 
The second question was whether these Euler norms bound the growth of the equivalent norms for the $\nu\geq3.125$e-5 Navier-Stokes cases.  The Euler norms do bound the norms from the $\nu\geq3.125$e-5 cases, which is why the additional $2e-6\leq\nu\leq1.5625$e-5 Navier-Stokes cases were run.

Could the Navier-Stokes $Z$, including the $\nu\leq1.5625$e-5 cases, be bounded by strong, 
but finite, Euler enstrophy growth?  If so, this would make it less likely that a 
$\nu\rightarrow0$ {\it dissipation anomaly} could  form at $t\approx 2t_x$. 
Or can the enstrophy continue to grow as $\nu$ decreases such that $Z\rightarrow\infty$ 
as $\nu\rightarrow0$?  First as $Z\sim 1/(\sqrt{\nu}B_x^2)$ at $t=t_x$ and 
then as $Z\sim \epsilon\nu^{-1}$ at $t_\epsilon\approx 2t_x$ as in figure \ref{fig:QsnuZ} 

{\bf Euler results} Four Euler cases were run, three in a $\ell=\pi$ domain with 
resolutions of $512^3$, $1024^3$ and $2048^3$, and a fourth case in a $(9\pi)^3$ domain 
using $2048^3$ grid points.  
Figure \ref{fig:EulerNSomaxtsmall} compares the Euler and Navier-Stokes 
$\sbl{\omega}{\infty}{}$. 

While the $\sbl{\omega}{\infty}{}$ for the three $\ell=4\pi$ Euler calculations agree 
up to $t=18$, after $t=18$ the $\sbl{\omega}{\infty}{}$ increase as the resolution is improved 
in a manner 
consistent with the exponential of exponential growth found for anti-parallel interactions 
\citep{Kerr2013b}.  The effect of the domain size upon the Euler norms is indicated by
the $(9\pi)^3$, $2048^3$ calculation, with its $\sbl{\omega}{\infty}{}$ closely following 
$\sbl{\omega}{\infty}{}$ from the $1024^3$ $(4\pi)^3$ case, 
two calculations with roughly the same local resolution. Therefore, using a $(4\pi)^3$ is sufficient
for determining Euler regularity properties, implying that for $\sbl{\omega}{\infty}{}$,
the $(4\pi)^3$, $2048^3$ calculation that is resolved until $t=24$ is the best for comparing
to all the Navier-Stokes calculations.

{\bf Navier-Stokes comparisons.} 
Next, we want to compare $\|\omega(t)\|_\infty$ from the very low viscosity, early time 
Navier-Stokes calculations to the $(4\pi)^3$, $2048^3$, Euler values. Up to $t=24$ for
the two resolved Navier-Stokes calculations, and even for the $\nu=1.5625$e-5 case
up to $t=19$, the $\|\omega(t)\|_\infty$ are bounded by the Euler values.  This suggests that 
the superexponential, non-singular growth of $\|\omega\|_\infty$ from the $\ell=4\pi$
Euler evolution should bound $\|\omega\|_\infty$ for all the viscous cases.  Preliminary tests 
using higher-order $\sbl{u}{\dot{H}^s(\bbT_\ell^3)}{}$ show that the 
$s=2$ and $s=5/2$ Euler norms are also bounding their Navier-Stokes counterparts.

However, the situation is different for the scaled enstrophies $(t_x-t)Z$ 
in figure \ref{fig:EulerNS-Zttx}.  The Navier-Stokes values 
do exceed the Euler values over an increasing timespan in a consistent manner as $\nu$ 
decreases, especially for $\nu\leq3.125$e-5. 
The scaling with an extra factor of $(t_x-t)$ helps highlight these differences 
in the Navier-Stokes enstrophy growth at early times. 
For the lowest viscosity case, $\nu=2$e-6 in a $(12\pi)^3$ domain, 
its $(t_x-t)Z$ exceeds the Euler values until $t=21$ and in figures \ref{fig:QsnuZ}
and \ref{fig:QHdRisnuZtime}, the grey-dashed curve shows that 
this eventually connects to the convergence of $\sqrt{\nu}Z$ at $t=t_x\approx40$.  
$\sbl{u}{\dot{H}^s(\bbT_\ell^3)}{}$, for $s=3/2$ and $\nu\geq1.5625$e-5 shows similar, 
but much weaker, signs of exceeding its Euler values.

\section{Summary and unanswered questions \label{sec:summary}} 

The trefoil calculations and conclusions presented in this paper have relied upon two types 
of weaves.  One weave generates the perturbed trefoil vortices with a 
single, dominant initial reconnection.  

The second weave shows how restrictions upon the growth of enstrophy can be avoided so that
the new scaling regime given in figures \ref{fig:QsnuZ}, \ref{fig:QHdRisnuZtime} and 
\ref{fig:dNSsnuZ} can be identified. A regime that might lead to the formation of
a Navier-Stokes {\it dissipation anomaly} at later times. That is, $\nu\to0$ smooth solutions, that 
generate finite energy dissipation in a finite time without singularities 
or additional roughness terms, 

It was the long, temporal extent of the trefoils' $\sqrt{\nu}Z(t)$ growth that first 
indicated the existence of this new scaling regime, with the last step provided by
new anti-parallel calculations that provide a mechanism for how the $(\sqrt{\nu}Z(t))^{-1/2}$
scaling regime begins.
A useful way to summarise the avoidance steps is to first list them in the order found, 
then restate them in terms of the forwards-in-time steps of the evolution.

The following steps in the evolution of the trefoil's enstrophy have been identified,
\ENM\item First, as illustrated in figure \ref{fig:QsnuZ}a, 
it was noticed that for each core radius, all of the $\sqrt{\nu}Z(t)$ crossed
at a common time $t_x$ so long as the domain size $\ell$ increased appropriately as
the viscosity $\nu$ decreased. 
The $\nu=3.125$e-5 curves in different computational domains show why increasing $\ell$ is needed 
for maintaining the convergence of $\sqrt{\nu}Z(t)$ at $t=t_x$. 

\item[] The large-scale growth of $h<0$, negative helicity, in figure \ref{fig:T45} 
could be providing 
a mechanism that allows small-scale $h>0$ and enstrophy $Z$ to grow, a mechanism that can be 
suppressed if the periodic boundaries are too close.

\item Linearly decreasing $(\sqrt{\nu}Z(t))^{-1/2}=B_\nu(t)$ \eqref{eq:Bnu} 
for $t\lesssim t_x$ was then identified using figure \ref{fig:QsnuZ}c,
from which the $T_c(\nu)$ needed for the next step were identified. 

\item Using these $T_c(\nu)$ and $B_x$, the $t\leq t_x$ linear collapse regimes were identified 
by determining and plotting 
$(T_c(\nu)-t_x)\bigl(B_\nu(t)-B_x\bigr)$ \eqref{eq:isnuZBxtime} 
for both the Q-trefoil and anti-parallel calculations, as shown 
in figures \ref{fig:QHdRisnuZtime} and \ref{fig:dNSsnuZ} respectively.

\item What physical event determines the end of the linear $(\sqrt{\nu}Z(t))^{-1/2}$ collapse?
This is identified as the end of the first reconnection using figure \ref{fig:T45} and
figure \ref{fig:antiPt16-24} for the trefoil and anti-parallel respectively.

\item What determines the beginning of the linear $(\sqrt{\nu}Z(t))^{-1/2}$ collapse?
The anti-parallel reconnection calculations identify the beginning of the linear collapse 
as when the vortices first meet with a brief spurt of circulation exchange that 
converts a finite fraction of the original 
$\Gamma_y$ into the $\Gamma_z$ of the new vortices at $t\approx t_\Gamma$. 
\EEN

By reversing these steps the multi-step origins of the new reconnection
scaling are these: Inviscid evolution from $t=0$ until the vortices touch, 
from which a viscous $(\sqrt{\nu}Z(t))^{-1/2}$ scaling regime forms 
that lasts until the first reconnection ends at $t_x$. 
Figure \ref{fig:T45} at $t\!=\!45>t_x\!=\!40$ shows that the necessary increase in
the enstrophy is associated with local increases in positive helicity and to
preserve the global helicity ${\cal H}$, the $h\!>\!0$ increases are
balanced by the generation of $h\!<\!0$ at larger scales.  The $t\leq t_x$ phase is 
followed a slow decay of the global helicity ${\cal H}(t)$ and further growth in the enstrophy 
until a there is a finite rate of energy dissipation at $t=t_\epsilon$, 
as shown in figure \ref{fig:QsnuZ}b. 
As this continues, a {\it dissipation anomaly} forms.
That is, finite energy dissipation in a finite time is generated. 

{\bf Further constraints on growth.}
Let us consider the ways that this progression of increasing $Z$ could be disrupted, 
and the evidence that it is not.

While increasing the domain skirted around the \cite{Constantin86} restrictions, 
since Euler $\sbl{\omega}{\infty}{}$ seems to bound the growth of all Navier-Stokes vorticity 
maxima, could this lead to an upper bound upon the Navier-Stokes enstrophies $Z$?  
Especially since for Euler both $\sbl{\omega}{\infty}{}$ and 
$Z=\sbl{\omega}{H^0(\bbT_\ell)}{2}=\sbl{u}{H^1(\bbT_\ell)}{2}$  are bounded.

Figure \ref{fig:EulerNS-Zttx} addresses this possibility using early time results from
very small viscosity calculations with $\nu\leq1.5625$e-5. These show that
the growth of the Navier-Stokes enstrophies exceeds the growth of the Euler enstrophy 
in a manner that connects the early growth of $Z$ to the convergence of
$\sqrt{\nu}Z(t)$ at $t=t_x\approx40$ in figure \ref{fig:QsnuZ}. 
For these periodic calculations, the Navier-Stokes enstrophy growth rates are bounded by the 
$\sbl{\omega}{\infty}{}$ only in the sense that $Z(t)\leq \ell^3(\nu)\sbl{\omega}{\infty}{}$.

Despite these observations, is there mathematics that can identify 
lower bounds for $\nu_s(\infty)$ as $\ell\rightarrow\infty$? 
\cite{Constantin86} points out that there are whole space $\bbR^3$ versions of the $\bbT^3_\ell$ 
inequalities used to bound convergence of the Navier-Stokes solutions to the Euler solutions 
\eqref{eq:Hsbnd} that can  be used to show the existence of critical $\nu_s(\ell=\infty)$ ($\bbR^3$). 
Some implications of this have been discussed by \cite{Masmoudi2007} 
and the empirical evidence in section \ref{sec:regularity} is consistent 
with the existence of $\nu_s(\infty)$ in the sense that the Navier-Stokes 
$\sbl{\omega}{\infty}{}$ are bounded by the Euler $\sbl{\omega}{\infty}{}$, but not
in the sense that $\sbl{\nabla\times(u(t)-v(t))}{\infty}{}\to0$ as $\nu\to0$.

However, it should be pointed out that an explicit description of the inequalities needed
to extend the bounds in \cite{Constantin86} to $\bbR^3$ has never been given and should be
provided. And even if the
$s>5/2$ higher-order, whole space Sobolev norms $\sbl{u}{H^s(\bbR^3)}{}$ are bounded,
it is possible that the dependence of the embedding theorems for $Z=\sbl{u}{H^1(\bbR^3)}{2}$ 
upon $\nu^{-1}$ would still allow unbounded growth of $Z$ as $\nu\to0$. Growth that
could still allow finite $\epsilon=\nu Z$ to form.

{\bf Future work}
A topic for later work is to properly identify the large-scale dynamical mechanism that 
allows the distant boundaries to constrain the growth of enstrophy within the original envelope 
as $\nu$ decreases. Once this mechanism is known, we would not only understand how
this constraint can be relaxed by increasing $\ell$, the domain size, but also how the 
large length scales of a turbulent flow can interact with the small, dissipation scales. 
The negative helicity identified in the outer loop of figure \ref{fig:T45} could 
be important clue because the transport of negative helicity to the large scales would allow
the positive helicity within the trefoil to increase as the vorticity cascades to small
scales. 

Further evidence for the existence of large-scale negative helicity comes 
from preliminary analysis of helicity spectra that shows that the post-reconnection 
low wavenumbers, that is large physical scales, are dominated by ${\cal H}(k)<0$.
However, it will be a challenge to identify the associated tendrils of ${\cal H}(x)<0$ in the
outer reaches of the $(6\pi)^3$, $(9\pi)^3$ and maybe $(12\pi)^3$ simulation data and see
physically what blocks their continued growth, and how that affects the dynamics within the
trefoils.

Is the $(\sqrt{\nu}Z(t))^{-1/2}$ scaling regime unique to these isolated vortex reconnection 
events?  To demonstrate that this regime is not unique to these calculations, let us 
consider the $D_m$ hierarchy of rescaled higher-order vorticity moments identified 
by the nonlinear time inequality analysis of \cite{Gibbon2010}.
\EQL{eq:Dm} D_m(t)=\varpi^{-\alpha_m}\sbl{\omega}{L^{2m}}{\alpha_m}\quad{\rm with}
\varpi=\ell^2/\nu\quad{\rm and}\quad\alpha_m=2m/(4m-3)\EN
In \cite{Donzisetal2013} it was shown that
$D_m(t)\geq D_{m+1}(t)$ for several sets of simulations, including one of the earlier 
anti-parallel cases \citep{Kerr2013a} with the same initial condition as used here.
Because the $D_m(t)\geq D_{m+1}(t)$ 
was particularly strong exactly over the time span of the $(\sqrt{\nu}Z(t))^{-1/2}$ scaling 
regime in figure \ref{fig:dNSsnuZ},
this suggests investigating additional
turbulence data sets, including all those used in \cite{Donzisetal2013}, 
to determine whether the new $(\sqrt{\nu}Z(t))^{-1/2}$ scaling 
and the $D_m(t)\geq D_{m+1}(t)$ hierarchy are ubiquitious during strong reconnection events.

\section*{Acknowledgements}

This work was stimulated by a visit to the University of Chicago in November 2013 and
subsequent discussions with W. Irvine at several meetings. 
I wish to thank J.C. Robinson at Warwick for his assistance in identifying the bounds upon 
small viscosity limits for the Navier-Stokes equations and S. Schleimer for assistance in 
clarifying the meaning of writhe, twist and self-linking. This work has also benefitted from conversations
with H. K. Moffatt and others at the 2015 BAMC meeting, IUTAM events in Venice and Montreal in 2016 and
the 2016 PDEs in Fluid Mechanics workshop of the Warwick EPSRC 
Symposium on PDEs and their Applications.  Computing resources have been provided
by the Centre for Scientific Computing at the University of Warwick, including use of the
EPSRC funded Mid-Plus Consortium cluster.

\vspace{-6mm}

\end{document}